\documentclass[10pt]{ietbook}
\usepackage{graphicx,multirow}
\usepackage[vlined,linesnumbered,ruled,boxed]{algorithm2e}

 \usepackage{layout}
 \usepackage{url}
\usepackage{wrapfig}
\usepackage{bibentry}

\usepackage{tabularx} 
\usepackage{algorithmic}
\usepackage{amssymb}
\usepackage{bbold}

\nobibliography*

\usepackage{wrapfig,color,caption,subcaption,balance,multirow}

\usepackage{adjustbox,tabu}

\usepackage[vlined,linesnumbered,ruled,boxed]{algorithm2e}
 \usepackage{amsmath}

 \usepackage{caption}
  \usepackage{subcaption}

\usepackage{bibentry}

\makeatletter
\def\thickhline{%
  \noalign{\ifnum0=`}\fi\hrule \@height \thickarrayrulewidth \futurelet
   \reserved@a\@xthickhline}
\def\@xthickhline{\ifx\reserved@a\thickhline
               \vskip\doublerulesep
               \vskip-\thickarrayrulewidth
             \fi
      \ifnum0=`{\fi}}
\makeatother

\newlength{\thickarrayrulewidth}
\nobibliography*

 \usepackage{graphicx,multirow}

\usepackage[colorlinks, citecolor=blue]{hyperref}
\usepackage{diagbox}

\newcommand{\eg}{{\it e.g., }}
\newcommand{\etal}{{\it et~al. }}
\newcommand{\ie}{{\it i.e., }}
\newcommand{\cc}{{confidential computing\xspace}}
\newcommand{\etc}{{\it etc. }}

\newcommand{\comments}[1]{}
\usepackage{cleveref}
\newcommand\hl{\bgroup\markoverwith
  {\textcolor{yellow}{\rule[-.5ex]{2pt}{2.5ex}}}\ULon}

\def\BibTeX{{\rm B\kern-.05em{\sc i\kern-.025em b}\kern-.08em
    T\kern-.1667em\lower.7ex\hbox{E}\kern-.125emX}}

\begin{document}


\rhbooktitle{Book title}

\markboth{Running head verso book title}{Running head recto chapter title}

\cauthor{Sm Zobaed and Mohsen Amini Salehi}

\chapter{Confidential Computing across Edge-to-Cloud for Machine Learning: A Survey Study}

Confidential computing has gained prominence due to the escalating volume of data-driven applications (\eg machine learning and big data) and the acute desire for secure processing of sensitive data, particularly, across distributed environments, such as edge-to-cloud continuum. Provided that the works accomplished in this emerging area are scattered across various research fields, this paper aims at surveying the fundamental concepts, and cutting-edge software and hardware solutions developed for confidential computing using trusted execution environments, homomorphic encryption, and secure enclaves. We underscore the significance of building trust in both hardware and software levels and delve into their applications particularly for machine learning (ML) applications. While substantial progress has been made, there are some barely-explored areas that need extra attention from the researchers and practitioners in the community to improve confidentiality aspects, develop more robust attestation mechanisms, and to address vulnerabilities of the existing trusted execution environments. Providing a comprehensive taxonomy of the confidential computing landscape, this survey enables researchers to advance this field to ultimately ensure the secure processing of users' sensitive data across a multitude of applications and computing tiers.


\section{Introduction}\label{sec:intro}


Every second, on average, more than 1.7 Megabytes of data is generated per person via different digital means including IoT sensors, organizational documentation, RSS feeds, transaction records, streaming, social media activities, and more to be processed by smart applications. The volume of data is growing faster than ever before, and it is expected that the world’s total data volume will exceed 149 zettabytes within the next three years~\cite{exa}. 
In response, hyperscaler cloud providers (\eg AWS, Azure, and Google cloud) offer various services for large-scale data storage, processing, and analysis in a distributed manner---across edge-to-cloud continuum. Although the wide adoption of AI/ML-based applications operating across IoT and edge-cloud systems has accelerated advancements in various aspects of human's life, they have created a larger data exposure surface, and less user-control over his/her data and infrastructure. In particular, the emergence of off-premises ML-based and generative AI (\eg ChatGPT \cite{sanderson2023gpt}) applications in almost every domain---from personalized healthcare, search, and archives to finance, assistive technology, and social networks~\cite{cloudacci}---has further complicated the matter of confidentiality and has brought it to the forefront of users' priorities. Despite various security services offered by cloud providers (\eg AWS IAM \cite{awsiam}, GuardDuty \cite{awsguard} and Security Hub \cite{awssechub}), organizations and individuals still have major concerns due to \textbf{lack of a natively-designed confidentiality solution} that encompasses the entirety of the hardware, systems (IoT-edge-cloud middleware), and application ecosystem.

Maintaining data confidentiality while data is \emph{``at rest''} (\eg stored on the cloud storage) is a widely known problem with well-studied solutions~\cite{zobaed2023ai}. Similarly, maintaining data confidentiality while it is \emph{``in transit''}) state has established solutions~\cite{koutsopoulos2017automated}, such as applying transport layer security (TLS) protocol~\cite{tls}. Nevertheless, the main data security and confidentiality challenge brought up by pervasive computing across IoT-edge-cloud is while the data is \emph{``in use''}, \ie while data is being processed \cite{datainuse}. As such, securing data-processing, particularly under distributed settings that has more exposure and are often controlled by multiple autonomous entities, has become the pressing desire. 
It is to address this very desire that the area of \textbf{\cc}, a.k.a. \textit{trusted computing}, is fast-emerging.

\begin{figure}
\centering
\includegraphics[width=.98\linewidth]{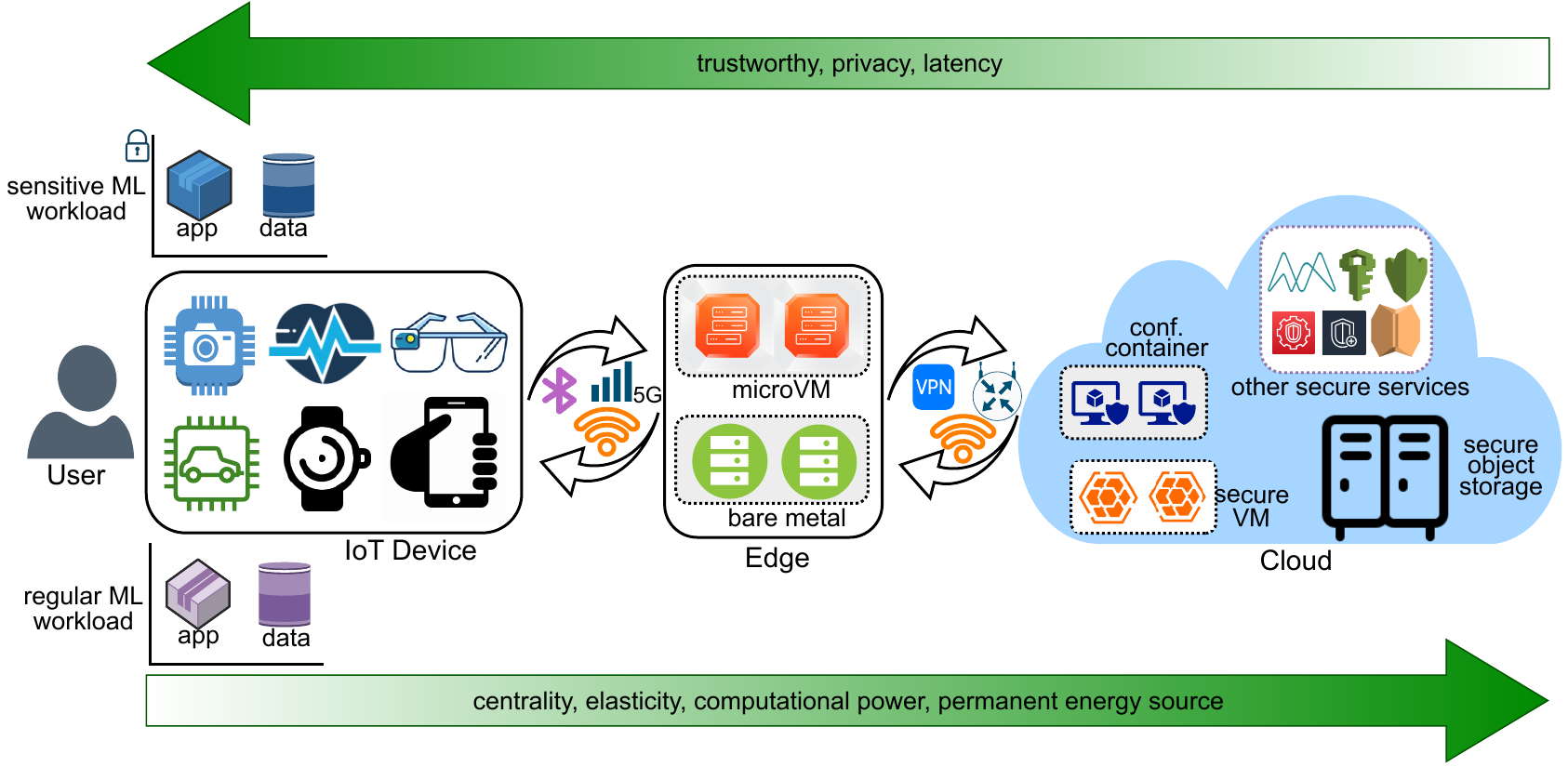}
\caption{Overview of confidential computing across edge-to-cloud continuum. Sensitive and regular ML workloads are requested from various sensors and IoT devices. Depending on the workload type, requests travel through various LAN and WAN networks: a LAN between the device and edge tiers (\eg via Bluetooth, Wi-Fi, and 5G); and a WAN between edge and cloud tiers (\eg via VPN, Wi-Fi, 5G, or a hybrid network). As a result, the workload can be executed collaboratively across secure services hosted on the IoT, edge, and cloud tiers, such as secure VM, container, and object storage; plus other secure cloud services  (\eg AWS Macie, IAM, KMS, GuardDuty, WAF, Shield).}
\label{fig:conf_high}
\end{figure}

With the prevalence of cloud computing, confidential computing over privacy-preserving data has to occur on shared pay-per-use infrastructures. Specifically, cloud services, such as Infrastructure as a Service (IaaS), traditionally offer computing infrastructure within Virtual Machines (VMs). While isolation and virtualization techniques (\eg VMs and containers) have been instrumental, the proliferation of cloud services has exposed software systems to new security vulnerabilities and increased risks~\cite{cloudacci, mulligan2021confidential} that are mainly rooted in the off-premise nature of clouds. To mitigate the threats introduced by the cloud services, edge-cloud systems have emerged not only to reduce the latency of accessing compute service, but also to enable maintaining sensitive user data on-premise (\ie at the edge)~\cite{ning2018preliminary,yu2017survey,gong2020intelligent}. However, as indicated in Figure~\ref{fig:conf_high}, there are two inherent problems with the edge computing paradigm: (\textbf{a}) They are resource- and (sometimes) energy-limited, hence, they are often insufficient to process massive volume of sensitive data within real-time constraints; (\textbf{b}) Despite of being on-premise and more trustworthy, edge nodes have their own vulnerabilities, due to wide dispersion and physical exposure. 

Although numerous studies have been undertaken to improve the real-timeness  and scalability across the edge-to-cloud, comparatively less attention has been paid to the confidential computing across heterogeneous tiers of the continuum. Figure~\ref{fig:conf_high} represents the path for confidential computing across the user/IoT devices to the edge, and cloud. The sensitive and regular data, generated by the devices, are first pre-processed on the devices using trusted applications; Then, it travels through various forms of LAN and WAN networks (\eg Bluetooth, Wi-Fi, VPN, and 5G) to be collaboratively processed by the services hosted on the edge and cloud tiers.

Establishing \cc\; of smart (often ML-based) applications within the context of IoT-edge-cloud entails understanding and enabling trust across hardware and software levels. At the hardware level, trust should be achieved via implementing proper root-of-trust flows that ensure a system is booted from a trusted image (signed by a trusted provider) into a valid state. Hardware vendors, such as Intel, AMD, and ARM, have come forward with Trusted Execution Environment (TEE) \cite{shepherd2016secure} technology that provides a hardware-assisted fully isolated execution environment that is deemed as the core of \cc. TEE offers a tamper-resistant processing environment, running in a separate kernel providing an acceptable level of originality, integrity, and confidentiality for the data and code to be executed~\cite{shepherd2016secure,sabt2015trusted,ning2018preliminary}.
\textit{The advantage of leveraging TEE-enabled cloud and edge infrastructures is that these systems remain secure even if their system software are compromised}. According to the Confidential Computing Consortium (CCC), TEE is the foundation for \cc, however, any TEE solution should also provide the ability for remote attestation, so that its trustworthiness can be verified. An application is considered as trusted, if it is executed within the TEE, its source code is available and inspectable, built into binaries with bit-exact reproducibility, and signed by a trusted authority~\cite{mulligan2021confidential}. Moreover, the data stored on and processed by a trusted application must be protected, and every interaction with any untrusted segments (\ie inter-application and remote service calls) is executed securely.

Provided the increasing prevalence and prominence of \cc\; across IoT-edge-cloud systems, and their large exposure surface, \textit{the \textbf{goal} of this study is to pinpoint the scope and challenges of \cc\; within the context of edge-cloud systems}. For that purpose, this study includes the following contributions:
\begin{enumerate}
    \item Describing the nuts and bolts of \cc\; across edge-cloud (Section~\ref{sec:keycomp}). 
\item Presenting a comprehensive taxonomy that shows the scope of \cc\; across edge-to-cloud (Section~\ref{sec:tax}).
\item Demystifying the anatomy of the \textit{trusted hardware technologies} as the main prerequisite to establish \cc\; across the continuum (Section~\ref{conf com: hardw}). 
\item Surveying the spectrum of middleware solutions for \cc\; that spans from the user-device level to the cloud data centers (Section~\ref{sec:middleware}). 
\item Elaborating on the trusted application development frameworks to build confidential software solutions, particularly, for ML  (Section~\ref{sec:application}). 
\item The last contribution of this study is in Section~\ref{sec:conclsn} where we identify the areas of research that demand further exploration from the community to overcome major remaining challenges of \cc\; across edge-to-cloud.
\end{enumerate}


\section{Key Concepts of Confidential Computing}
\label{sec:keycomp}

\subsection{Trusted Execution Environment (TEE)}
\begin{figure}
\centering
\includegraphics[width=.55\linewidth]{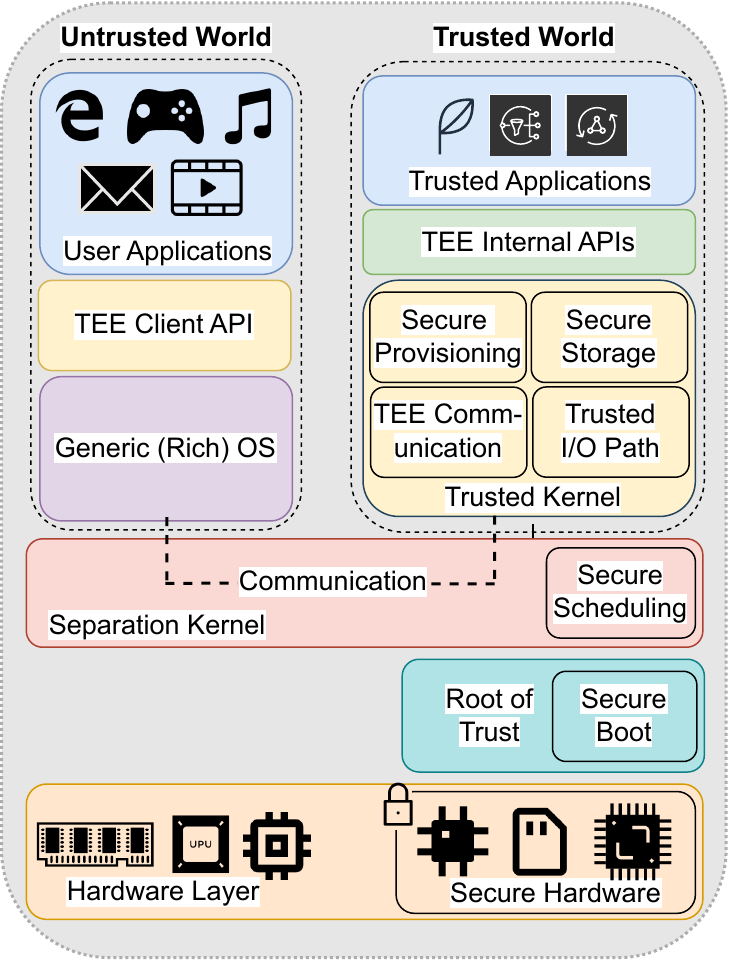}
\caption{High-level architectural overview of TEE building blocks}
\label{fig:tee_archi}
\end{figure}

A trusted execution environment (TEE) is a tamper-resistant processing environment that executes on a separation
kernel. In confidential computing, TEEs guarantee the protection of the code and data loaded inside an isolated area, which is also referred to as \emph{enclave}. The design goal of TEE is to prevent the manipulation of various software adversaries (\eg malware and hacked OS) or even hardware adversaries who have physical access to the platform.
TEE offers hardware-enforced security features that include isolated execution, guaranteeing the integrity and confidentiality of the enclave, and the ability of code authentication to execute inside
the enclave through attestation. Figure~\ref{fig:tee_archi} represents the building blocks and the core design aspect of TEE and we shed light on them in the rest of this section.


\subsection{Building Blocks of TEE}
GlobalPlatform is a widely known international association that identifies, develops, and promotes technical standards to facilitate secure and interoperable deployment and management of embedded applications on secure chips. They have released the detailed hardware and software architecture of TEE highlighting full isolation between Rich OS Environment (ROE) and TEE~\cite{shepherd2016secure}. 
Each environment separately allocates their own resources \eg RAM, ROM, CPU, and OS, while communications between the
two environments are only performed via the TEE Client
API. 

\textbf{{Kernel Separation}} is a fundamental concept of the TEE architecture, as it underpins the essential property of isolated execution. This concept was first introduced in the context of a dual-execution environment model \cite{sabt2015dual}. The primary objective of the separation kernel is to facilitate the cohabitation of diverse systems requiring different security levels on a single platform. The separation kernel divides the system into several partitions, ensuring robust isolation between them, except for the provision of a controlled interface for inter-partition communication. To assess the credibility of a kernel separation implementation, Sabt \etal developed a trust function based on its integrity measurement metrics that returns the level of trust of a given TEE~\cite{sabt2015trusted}.

The specific security requirements for separation kernels are outlined in the Separation Kernel Protection Profile (SKPP) \cite{cinque2022virtualizing}. SKPP describes the separation kernel as a combination of hardware, firmware, and/or software mechanisms with a core function to establish, isolate, and manage information flow between the subjects and the partitions. In contrast to conventional security kernels like operating systems, micro-kernels, and hypervisors, the separation kernel is more streamlined, offering both time and space partitioning, thus, contributing to system security and efficiency.

\textbf{{Secure Boot}} is a key component of a Trusted Execution Environment (TEE) that assures only authenticated software can be executed during the system boot-up phase. This mechanism acts as a critical line of defense against tampering attempts on boot loaders, key operating system files, and more. A device equipped with secure boot can interrupt the bootstrap process if modifications in the loaded code are detected, reinforcing the integrity of the system's operations.

Implemented at the firmware level, secure boot operates on the principle of trust chain establishment. As illustrated by Arbaugh \etal \cite{arbaugh1997secure}, secure boot verifies the integrity of each subsequent component according to a predefined reference value. This process unfolds across several stages, forming a chain of trust. The starting point of this chain is the Root of Trust (RoT), typically a hardware-based module, which is intrinsically trusted to form the foundation for verifying the trustworthiness of subsequent layers.
Upon system start-up, the RoT initiates the process by verifying the bootloader's integrity. Once validated, the bootloader checks the integrity of the next software component, and this sequence continues. Each link in the chain is responsible to validate the subsequent component's authenticity and integrity before allowing its execution. This methodical and layered approach to validation ensures that the execution environment remains secure and trustworthy.

A critical part of secure boot in TEEs involves the use of cryptographic signatures. Each software component involved in the boot process is signed using a private key. During the boot process, the signature of each component is verified against a trusted public key, certifying the software's origin and maintaining its integrity.

\textbf{{Secure Scheduling}}
assures a \emph{balanced} and \emph{efficient}
coordination between the TEE and the rest of the system.
Indeed, it should assure that the tasks running in the TEE do not affect the responsiveness of the main OS. Thus, the scheduler is often designed preemptive. Furthermore, the scheduler should take real-time constraints into consideration.
Authors in~\cite{sangorrin2012integrated} propose a secure scheduler that enhances the
responsiveness of the main OS without compromising the real-
time performance of the system.

\textbf{{Inter-Environment Communication}} is the mechanism that allows the TEE to interact with the rest of the system. While this provides numerous benefits, it also presents new threats, including message overload attacks, user and control data corruption attacks, memory faults caused by shared page removals, and unbound waits due to the non-cooperation of the system's untrusted part \cite{sabt2015trusted}.

The chosen mechanism for inter-environment communication must ensure reliability, maintain minimal overhead, and protect communication structures. Several models of communication have been identified in the literature, including the GlobalPlatform TEE Client API, the secure Remote Procedure Call (RPC) of Trusted Language Runtime~\cite{santos2014using}, and the real-time RPC of SafeG \cite{sangorrin2013reliable}. These models are designed to address the challenges associated with secure inter-environment communication and contribute to the secure and efficient operation of the TEE.

\textbf{{Secure Storage}} is a fundamental component of a TEE, designed to ensure the confidentiality, integrity, and freshness of stored data, thereby preventing replay attacks and maintaining state continuity \cite{jangid2021towards}. Access to this data is tightly controlled which ensures only authorized entities can access it \cite{sabt2015trusted}. A popular implementation approach is through sealed storage. This strategy is predicated on three core components: (1) an integrity-protected secret key exclusively accessible by the TEE; (2) cryptographic protocols, including authenticated encryption algorithms; and (3) a mechanism for data rollback protection, such as replay-protected memory blocks (RPMB) \cite{li2019research}.

\textbf{{Trusted I/O Path}} is designed to safeguard the authenticity and, if necessary, the confidentiality of the communication channel between the TEE and peripherals such as keyboards or sensors. This design strategy ensures that input and output data are protected from eavesdropping or tampering by malevolent applications. Specifically, the Trusted I/O path provides protection against four major classes of attacks: screen-capture, keylogging, overlaying, and phishing. The establishment of a trusted path to user-interface (UI) enabled devices extends the functionality within the TEE, facilitating direct interaction between the human user and applications running within the TEE.

\subsection{Measuring the Trustworthiness of a System}
Evaluating trust, especially in a comparative sense between different systems employing TEEs, requires a mechanism to quantify trust. In colloquial language, trust refers to the conviction of the integrity and reliability of a person or entity. This subjective property, while intrinsic to human relationships, is challenging to numerically capture. In computer systems, the notion of trust becomes even more nuanced. Here, an entity is typically trusted if it has consistently behaved as expected and will continue to do so in the future.

In the realm of computing, trust can be categorized as static or dynamic. Static trust is derived from an exhaustive evaluation against specific security benchmarks before the deployment of a system. A prominent example of a standard providing assurance measures for security evaluation is the Common Criteria~\cite{sabt2015trusted}. This standard enumerates seven evaluation assurance levels (EAL1-EAL7), with each higher level encompassing the requirements of its predecessor. Thus, under static trust, the trustworthiness of a system is assessed once before its deployment.
Conversely, dynamic trust is contingent upon the state of an active system and fluctuates accordingly. With the trust status of a system undergoing continuous changes, dynamic trust necessitates a regular assessment of the system's trustworthiness throughout its lifecycle.

TEE systems embody a hybrid form of trust – encompassing both static and semi-dynamic aspects. Prior to deployment, a TEE is subjected to stringent certification procedures, including verification of its security level according to a protection profile, which outlines a predefined set of security requirements. For instance, the GlobalPlatform protection profile conforms to EAL2~\cite{sabt2015trusted}. Moreover, each boot sequence includes a  `Root of Trust' (RoT) check, ensuring that the TEE in operation is the certified one provided by the platform vendor. This safeguards the integrity of the TEE code~\cite{fei2021security,mo2022sok}.

Once operational, the TEE code's integrity is protected by the underlying separation kernel. As such, the trust in TEEs is considered semi-dynamic – the TEE is not expected to change its trust level while running due to the protection provided by the separation kernel. In this trust model, the trust measurements are integrity assessments, and the trust score is a binary indicator of the code's integrity state. The TEE is deemed trusted when its trust score is true and untrusted otherwise. The reliability of this trust score hinges on the integrity measurement definitions.

To assess the real trust value, Sabt \etal proposed a trust function $f$(TEE, protection profile, RoT, measurements)~\cite{sabt2015trusted}. This function returns the trust level of a given TEE based on three parameters: the certifying protection profile, the reliability of RoT, and the integrity measurements. Such a function provides a quantitative basis to weigh trust in TEE systems.

\subsection{Trustworthy Code Execution}
Confidential computing fundamentally ensures trustworthy code execution through the strategic isolation of enclaves, or Trusted Execution Environments (TEEs), from untrusted environments \cite{fei2021security}. This secure partitioning is facilitated by the rigorous protection and management of enclave memory sections, orchestrated via the hardware and system software of the TEE. Distinct processors such as ARM, Intel, and AMD have each conceived their unique architectural approaches to facilitate this isolation.

Intel Software Guard Extensions (SGX) \cite{chen2019sgxpectre} employs a memory encryption engine to provide assured secrecy, integrity, and freshness of the CPU-DRAM traffic within the enclave memory ranges. In contrast, ARM TrustZone \cite{valadares2021trusted} utilizes distinct page tables, hardware privilege layers (\eg EL3 and Secure EL2/EL1/EL0), and a TrustZone address space controller (TZASC4) for its implementation.


\subsection{Remote Attestation in Confidential Computing}
Remote attestation~\cite{fei2021security,menetrey2022exploratory} enables the user (a.k.a. confidential workload owner) to ascertain the level of trust/integrity of a remote TEE prior to transmitting sensitive data/code. It enables the owner to authenticate the hardware, validate the trusted state of a distant TEE, and determine whether the intended program is operating securely within the TEE. A third party, besides the user and host, could perform the attestation~\cite{menetrey2022exploratory,mo2022sok}. Also, the service provider and enclave owner could execute the attestation directly~\cite{mo2022sok}. An attestation server could be launched by the processor manufacturer, as with Intel's SGX attestation service~\cite{chen2019sgxpectre}, which is distinct from the cloud service provider.

\section{Taxonomy of Confidential Computing across Edge-to-Cloud}
\label{sec:tax}
\begin{figure} [!htbp]
\centering
\includegraphics[width=.99\linewidth]{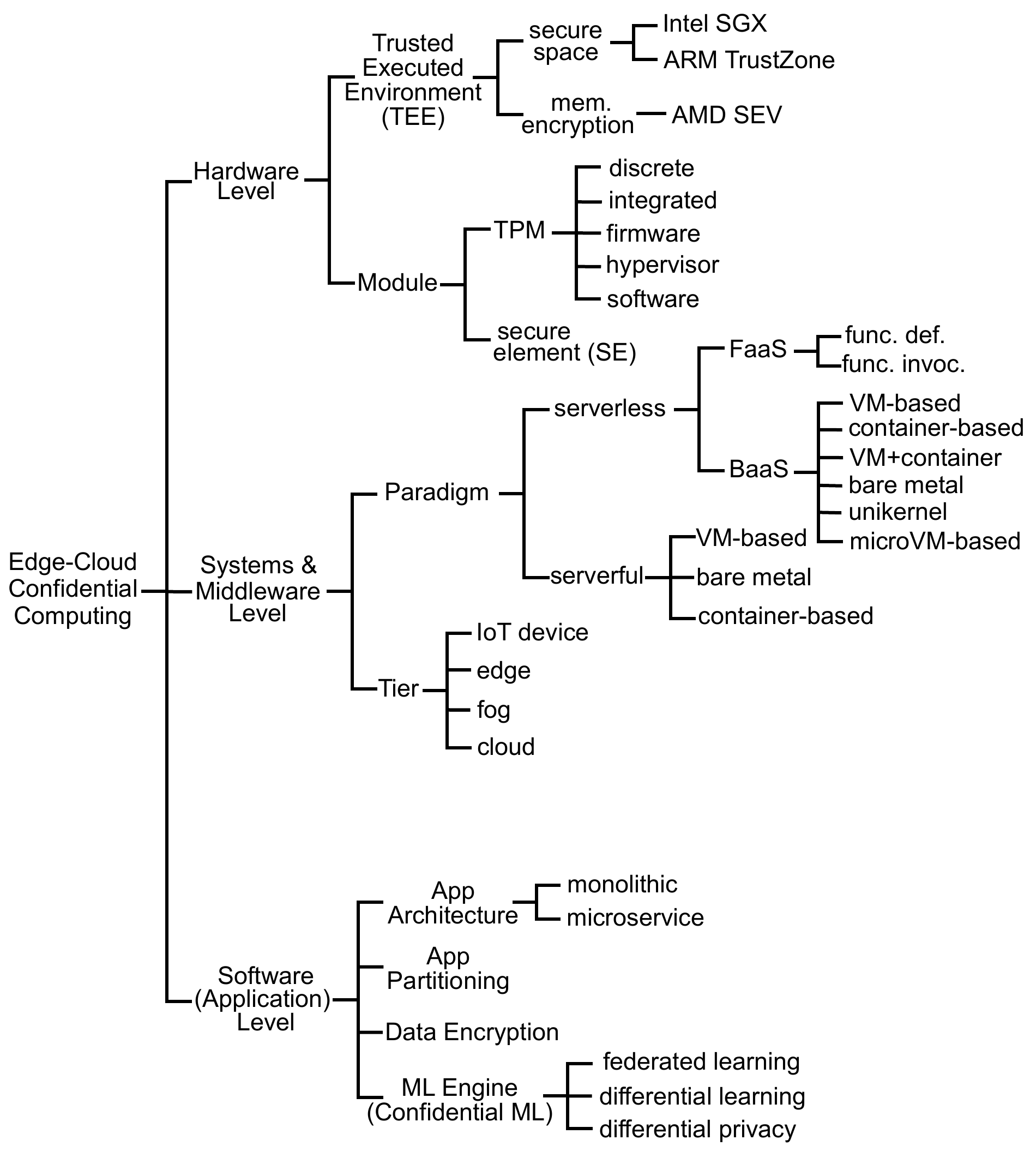}
\caption{Taxonomy of confidential computing encompassing hardware level, systems and middleware, and software application levels}
\label{fig:taxonomy}
\end{figure}

Figure~\ref{fig:taxonomy} represents a comprehensive taxonomy of the scopes where confidential computing can be implemented, particularly, with respect to edge-to-cloud continuum and machine learning (ML) applications. The first level of the taxonomy broadly covers confidential computing at the hardware, system middleware, and application levels. These three are the main thrusts of this paper and subsequent Sections~\ref{conf com: hardw} to \ref{sec:application} essentially elaborate on each one of these thrusts.

Hardware is the cornerstone of the ongoing momentum towards confidential computing. At the \textit{hardware} level, confidential computing is achieved in two main ways: (A) \textit{Trusted Execution Environments (TEE)} \cite{sabt2015trusted} that deals with the secure execution of applications; and (B) \textit{Secure Hardware Modules}~\cite{securehw} that are not in charge of application execution and deal with other aspects such as secure storage, secure key management, \etc. Prominent implementations of TEE, namely Intel Software Guard Extension (SGX) \cite{weiser2017sgxio}, AMD Secure Encrypted Virtualization (SEV) \cite{AmdSev}, and ARM TrustZone \cite{valadares2021trusted}, are studied in Section~\ref{teehardware}. The other hardware category, \textit{Secure Modules}, includes two main classes: Trusted Platform Module (TPM), and Secure Element (SE) that are elaborated in Section~\ref{hwmodules}. 

At the system software (a.k.a. \textit{middleware}) level, we have considered the \textit{computing paradigm} and \textit{computing tier} aspects where the former encompasses serverless and serverful computing paradigms. 
The \textit{computing tier} deals with the device, edge, fog, and cloud tiers and studies the confidential computing research works undertaken within and across these tiers.

At the \textit{application level}, we discuss how developers and programming-level tools can reinforce or undermine confidential computing. In particular, we study the impact of software architecture (monolithic vs micro-service) used to develop an application on confidential computing. We unfold the structure of monolithic confidential computing applications (\eg S3C~\cite{woodworth2016s3c}, S3BD~\cite{woodworth2019s3bd}, and ClusPr~\cite{zobaed2022privacy}), as well as microservice-based ones (\eg SAED~\cite{zobaed2021saed}). What is more, we study how application partitioning (across edge-cloud) and application-data encryption can influence confidential computing. 

Last but not least, at the application level, in Section~\ref{confml}, we pay specific attention to the interplay of machine learning (ML) and confidential computing that have given birth to a new research field, called \textit{confidential ML}. In particular, we study federated learning, differential learning, and differential privacy techniques and their mutual impact with the confidential computing.

Confidential computing research is not a monolithic field, but rather it is a rich tapestry of interconnected research areas. Each of these areas contribute to our understanding and ability to implement confidential computing. To illuminate the breadth and depth of this field, we have compiled a comprehensive summary of prior studies in Table~\ref{tab:survey}. This table categorizes each study according to its alignment with one or more areas of our taxonomy. By viewing these studies in aggregate, we can discern correlations, contributions, and gaps in the existing research. A more thorough view to the research works listed in the table provides a deeper understanding of the state-of-the-art in confidential computing. The table columns are in alignment with the categorization introduced in the taxonomy of Figure~\ref{fig:taxonomy}. In particular, for each study, we have highlighted the type of hardware being used and the computing paradigm being targeted by that study. The ``Remarks'' column focuses on the shortcoming of that study and highlights the potential future studies that can be accomplished.
\begin{table*}[]
\caption{Prior studies undertaken on different aspects of edge-cloud confidential computing}
\label{tab:survey}
\resizebox{\linewidth}{!}{
\begin{tabular}{|l|l|l|l|l|l|}
\hline
\textbf{Prior Studies}                                  & \textbf{Hardware} & \textbf{\begin{tabular}[c]{@{}l@{}}Computing \\ Paradigm\end{tabular}} & \textbf{\begin{tabular}[c]{@{}l@{}}Service\\ Model\end{tabular}} & \textbf{Summary} & \textbf{Remarks} \\ \hline
                                                            \hline
\multirow{2}{*}{\begin{tabular}[c]{@{}l@{}}Sabt \etal \cite{sabt2015trusted} \\(2015) \end{tabular}}              & \multicolumn{1}{l|}{\multirow{2}{*}{TEE}} & \multicolumn{1}{l|}{\multirow{2}{*}{User-end}}           & \multirow{2}{*}{IaaS}                                                    & \begin{tabular}[c]{@{}l@{}} + Present core components\\ of TEE \end{tabular}                                                                                                                                                   & - Study ARM TrustZone only                                                                                                                       \\ 
                                                   & \multicolumn{1}{l|}{}                     & \multicolumn{1}{l|}{}                            &                                                                      & \begin{tabular}[c]{@{}l@{}} + Survey  industrial \& academic\\TEE\end{tabular}                                                                                                                                            & - Discuss deprecated TEEs                                                                                                                             \\ \thickhline
\multirow{2}{*}{\begin{tabular}[c]{@{}l@{}}Shepherd \etal \\\cite{shepherd2016secure} (2016) \end{tabular}}      & \multicolumn{1}{l|}{TEE}                  & \multicolumn{1}{l|}{\multirow{2}{*}{User-end}}           & \multirow{2}{*}{\begin{tabular}[c]{@{}l@{}}Bare\\ Metal\end{tabular}}                                                       & \multirow{2}{*}{\begin{tabular}[c]{@{}l@{}} + Survey on versatile trusted\\ hardware components \end{tabular}}                                                                                                                 & {\begin{tabular}[c]{@{}l@{}} - Limited discussion on\\ hardwares' vulnerabilities    \end{tabular}}                                                                                                    \\ 
                                                   & \multicolumn{1}{l|}{TPM, SE}              & \multicolumn{1}{l|}{}                            &                                                                      &                                                                                                                                                                                    & \begin{tabular}[c]{@{}l@{}}- Little discussion on middle-\\wares and application framework\end{tabular}                                               \\ \thickhline

                                                   
\multirow{2}{*}{\begin{tabular}[c]{@{}l@{}}Ning  \etal \cite{ning2018preliminary} \\ (2018) \end{tabular}} & \multicolumn{1}{l|}{\multirow{2}{*}{TEE}} & \multicolumn{1}{l|}{\multirow{2}{*}{Serverful}}           & \multirow{2}{*}{IaaS}                                          & {\begin{tabular}[c]{@{}l@{}} + Evaluate edge-centric\\ TEE architectures\end{tabular}}                                                                                                                               & \begin{tabular}[c]{@{}l@{}} - Do not discuss SGX-based\\ TEEs\end{tabular}                   \\ 
                                                   & \multicolumn{1}{l|}{}                     & \multicolumn{1}{l|}{}                            &                                                                      & \begin{tabular}[c]{@{}l@{}} + Deploy TrustZone-based\\ TEE on edge                    \end{tabular}                                                                                                                                       & {\begin{tabular}[c]{@{}l@{}} - Do not discuss end-to-end\\ confidential computing    \end{tabular}}                                                                                                 \\ \thickhline

\multirow{2}{*}{\begin{tabular}[c]{@{}l@{}}Aublin~\etal\\\cite{Aublin2018libseal}  (2018) \end{tabular}}  & \multicolumn{1}{l|}{\multirow{2}{*}{TEE}} & \multicolumn{1}{l|}{\multirow{2}{*}{User-end}}           & \multirow{2}{*}{SaaS}                                          & \begin{tabular}[c]{@{}l@{}} \\ + Propose trusted logger service\\ library for TA \end{tabular}                                                           & - Focus on Intel SGX only                                                                                                                  \\ 
                                                   & \multicolumn{1}{l|}{}                     & \multicolumn{1}{l|}{}                            &                                                                      &                                                                                                                                                                                    & {\begin{tabular}[l]{@{}l@{}} - Do not guarantee end-to-end\\ confidential computing \end{tabular}}                 \\ \thickhline

\multirow{3}{*}{\begin{tabular}[c]{@{}l@{}}Nguyen~\etal\\\cite{nguyen2018logsafe} (2018)\end{tabular}}  & \multicolumn{1}{l|}{\multirow{3}{*}{TEE}} & \multicolumn{1}{l|}{\multirow{3}{*}{Serverful}}           & \multirow{3}{*}{SaaS}                                          & \begin{tabular}[l]{@{}l@{}} + Propose storing mechanism\\ for  IoT logs on SGX-enhanced \\ cloud nodes\end{tabular}                                                           & - Focus on Intel SGX only                                                                 \\ 
                                                   & \multicolumn{1}{l|}{}                     & \multicolumn{1}{l|}{}                            &                                                                      &                                                                                                                                                                                    & {\begin{tabular}[c]{@{}l@{}}- Do not ensure end-to-end\\ data confidentiality\end{tabular}}                                                                                                      \\ \thickhline

\multirow{2}{*}{\begin{tabular}[c]{@{}l@{}}Valadares~\etal\\\cite{ valadares2018achieving} (2018)\end{tabular}}  & \multicolumn{1}{l|}{\multirow{2}{*}{TEE}} & \multicolumn{1}{l|}{\multirow{2}{*}{User-end}}           & \multirow{2}{*}{SaaS}                                          & \begin{tabular}[c]{@{}l@{}}+ An attestation mechanism\\ for preventing physical\\ attacks on IoT \end{tabular}                                                           & - Data integrity is not assured                                                                                                                      \\ 
                                                   & 
                                                   & 
                                                   &                                                                      & \begin{tabular}[l]{@{}l@{}} + Propose secure key\\ management through key\\ vault running in SGX\end{tabular}                                                                         & {\begin{tabular}[l]{@{}l@{}} - Fall short to detect \\compromised source\end{tabular}}                                                                                                             \\ \thickhline


\multirow{2}{*}{\begin{tabular}[c]{@{}l@{}}Ayoade~\etal\\\cite{ ayoade2019secure} (2019)\end{tabular}}  & \multicolumn{1}{l|}{\multirow{2}{*}{TEE}} & \multicolumn{1}{l|}{\multirow{2}{*}{User-end}}           & \multirow{2}{*}{SaaS}                                          & \begin{tabular}[l]{@{}l@{}}+ A secured data processing \\ framework with isolated trusted \\ and untrusted modules\end{tabular}                                                           & - Focus on Intel SGX only                                                                                                                  \\ 
                                                   & \multicolumn{1}{l|}{}                     & \multicolumn{1}{l|}{}                            &                                                                      &                                                                                                                                                                                    & {\begin{tabular}[c]{@{}l@{}}- Do not ensure end-to-end\\ confidential computing \end{tabular}}        
\\ \thickhline

\multirow{2}{*}{\begin{tabular}[c]{@{}l@{}}Pinto \etal \cite{ pinto2019demystifying} \\(2019)\end{tabular} }      & \multicolumn{1}{l|}{\multirow{2}{*}{N/A}}    & \multicolumn{1}{l|}{\multirow{2}{*}{User-end}}           & \multirow{2}{*}{IaaS}                                          & {\begin{tabular}[c]{@{}l@{}} + Survey on TEEs and \\ hardware-assisted virtualization\end{tabular}}                                                                                                                              & \multirow{2}{*} {\begin{tabular}[c]{@{}l@{}} - Only focus on ARM TrustZone \end{tabular} }                                                                                                   \\ 
                                                   & \multicolumn{1}{l|}{}                     & \multicolumn{1}{l|}{}                            &                                                                      & {\begin{tabular}[l]{@{}l@{}} + Discuss vulnerabilities \\ of trusted computing \end{tabular}}                                                                                                                                      &                                                                                                                                                     \\ \thickhline
\multirow{2}{*}{\begin{tabular}[c]{@{}l@{}}Van \etal \cite{ van2019tale}\\(2019)  \end{tabular} }                & \multicolumn{1}{l|}{\multirow{2}{*}{TEE}} & \multicolumn{1}{l|}{\multirow{2}{*}{Serverful}}  & \multirow{2}{*}{IaaS}                                                    & \begin{tabular}[c]{@{}l@{}} + Discover vulnerabilities\\ in all open source SDKs \\ used for enclave deployment\end{tabular}                                                  & \multirow{2}{*} {\begin{tabular}[c]{@{}l@{}} - Only focus on trusted hardware\end{tabular} }                                                                                                \\ 
                                                   & \multicolumn{1}{l|}{}                     & \multicolumn{1}{l|}{}                            &                                                                      & + Study few TEE hardwares                                                                                                                                                   &                                                                                                                                                     \\ \thickhline

{\begin{tabular}[c]{@{}l@{}}Brenner  \etal \\\cite{brenner2019trust} (2019)\end{tabular}}          & \multicolumn{1}{l|}{\multirow{2}{*}{TEE}} & \multicolumn{1}{l|}{\multirow{2}{*}{Serverless}} & \multirow{2}{*}{FaaS}                                          & \multirow{2}{*}{\begin{tabular}[c]{@{}l@{}} + Study library dependencies \\ for securing FaaS clouds\end{tabular}} & \begin{tabular}[c]{@{}l@{}} - Do not consider TrustZone\\ and SEV-enabled hardwares\end{tabular}                                                     \\ 
                                                   & \multicolumn{1}{l|}{}                     & \multicolumn{1}{l|}{}                            &                                                                      &                                                                                                                                                                                    & - Consider enclave is attack free                                                                                                                   \\ \thickhline
\multirow{2}{*}{\begin{tabular}[c]{@{}l@{}}Ibrahim \etal \\ \cite{ibrahim2019trusted} (2019) \end{tabular} }        & \multicolumn{1}{l|}{TEE}                  & \multicolumn{1}{l|}{\multirow{2}{*}{User-end}}           & \multirow{2}{*}{IaaS}                                          & + Compare different TPMs                                                                                                                                                   & {\begin{tabular}[c]{@{}l@{}} - Briefly discuss TPM-driven\\ application framework \end{tabular}}                                                                                                    \\ 
                                                   & \multicolumn{1}{l|}{TPM}                  & \multicolumn{1}{l|}{}                            &                                                                      & \begin{tabular}[c]{@{}l@{}} + Discuss secure migration\\ and cloning of VMs                 \end{tabular}                                                                                                                     & {\begin{tabular}[l]{@{}l@{}}- Do not discuss using  TPM\\ across edge-cloud     \end{tabular}}                                                                                               \\ \thickhline
{\begin{tabular}[c]{@{}l@{}}Aslanpour  \etal \\\cite{aslanpour2021serverless} (2021) \end{tabular}} & \multicolumn{1}{l|}{\multirow{2}{*}{TEE}} & \multirow{2}{*}{\begin{tabular}[c]{@{}l@{}}Serverless\\ edge\end{tabular}}  & \multirow{2}{*}{\begin{tabular}[c]{@{}l@{}}Bare\\ Metal\end{tabular}}                                          & + Study serverless edge                                                                                                                                            & \multirow{2}{*}{\begin{tabular}[c]{@{}l@{}} - Only focusing on middleware\\ and applications\end{tabular} }                                                                                        \\ 
                                                   & \multicolumn{1}{l|}{}                     & \multicolumn{1}{l|}{}                            &                                                                      & \begin{tabular}[c]{@{}l@{}} + Discuss scopes and challenges\\ of serverless edge computing \end{tabular}                                                                         &                                                                                                                                                     \\ \thickhline
{\begin{tabular}[c]{@{}l@{}}Valadares \etal \\\cite{ valadares2021trusted} (2021)\end {tabular}}                & \multicolumn{1}{l|}{TEE}                  & \multicolumn{1}{l|}{User-end}                            & IaaS                                                           & {\begin{tabular}[c]{@{}l@{}} + Survey on TEEs that are\\ used in edge-cloud systems \end{tabular}}                                                                                                                                 & {\begin{tabular}[c]{@{}l@{}} - Do not cover cross-system\\ application frameworks \end{tabular}}                                                                                                 \\ \thickhline
                                                   
{\begin{tabular}[c]{@{}l@{}}Menetrey \etal \\\cite{menetrey2022exploratory} (2022)                 \end{tabular}}& \multicolumn{1}{l|}{TEE}                  & \multicolumn{1}{l|}{Serverful}                   & IaaS                                                           &  \begin{tabular}[c]{@{}l@{}} + Study attestation \\mechanisms for TEEs  \end{tabular}                                                                                                                                           & {\begin{tabular}[c]{@{}l@{}} - Do not discuss end-to-end\\ confidential computing    \end{tabular}}                                                                                                    \\ \thickhline

\multirow{2}{*}{\begin{tabular}[c]{@{}l@{}}Li  \etal \cite{li2021securing} \\ (2022)\end{tabular}}                 & \multicolumn{1}{l|}{\multirow{2}{*}{TEE}} & \multicolumn{1}{l|}{\multirow{2}{*}{Serverless}} & \multirow{2}{*}{FaaS}                                          & \begin{tabular}[c]{@{}l@{}} + Survey on serverless\\ applications deployed\\ in SGX enclaves\end{tabular}                                                                            & \multirow{2}{*}{\begin{tabular}[c]{@{}l@{}} \\- Do not consider SEV and \\TrustZone\end{tabular}}                            \\ 
                                                   & \multicolumn{1}{l|}{}                     & \multicolumn{1}{l|}{}                            &                                                                      & \begin{tabular}[c]{@{}l@{}} + Propose SGX extension for\\ confidential serverless jobs\end{tabular}                                                            &                                                                                                                                                     \\ \thickhline

\end{tabular}
}
\end{table*}
\section{Confidential Computing Hardware}
\label{conf com: hardw}

\subsection{TEE Hardware Technologies for Confidential Computing}\label{teehardware}
\textit{Secure Space} and \textit{Memory Encryption} are crucial TEE hardware technologies that underpin the functioning of confidential computing systems. Understanding these technologies and how they are applied is key to grasping how hardware-based TEEs function.

Secure space, also referred to as isolated execution, is a hardware feature that allows the creation of private regions within the main memory. These regions, also known as enclaves or secure enclaves, function as separate execution environments where code and data can be securely processed, isolated from the rest of the system. This is to ensure that sensitive data is accessible only to authorized processes and is shielded from any potentially malicious software running on the same hardware platform. Technologies such as Intel SGX~\cite{costan2016intel} and ARM TrustZone~\cite{valadares2021trusted} utilize this concept, creating isolated secure spaces where sensitive code and data can operate in a controlled, confidential manner.

Memory encryption is a technology used to secure data at rest, in use, and in transit within and between the processor and memory. By encrypting the data within memory, this technology makes it incredibly difficult for an attacker to make sense of any data they might access, effectively preserving the confidentiality and integrity of the data. The encryption and decryption processes are carried out within the CPU, making the procedure transparent to applications, and do not require any modifications to the software. This technology is widely used in AMD's Secure Encrypted Virtualization (SEV)~\cite{AmdSev} where all the memory contents of a virtual machine are encrypted, ensuring the security of data from any possible external threats.

The secure space and memory encryption technologies are not mutually exclusive and can often be combined in various ways to provide multiple layers of security, enhancing the overall security posture of the system.

Table~\ref{tbl:hardware} compares the attributes of these three TEE hardware implementations from various aspects. In the rest of this section, we elaborate on these attributes for each hardware and compares them against each other. 

\subsubsection{Intel Software Guard eXtension (Intel SGX)} 
\begin{figure} [!htbp]
\centering
\includegraphics[width=.52\linewidth]{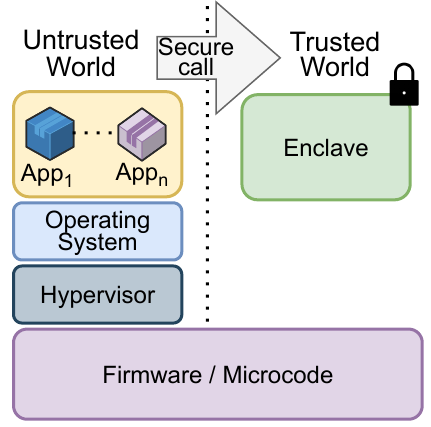}
\caption{Illustration of secure isolation provided by Intel SGX. SGX consists of enclaves that offer a trusted world where sensitive code and data are isolated from the rest of the system to maintain integrity and confidentiality.}
\label{fig:tee_sgx}
\end{figure}

Intel SGX, an extension to the x86-64 instruction set, provides a facility for executing select portions of an application within a hardware-assisted TEE, specifically within a secure enclave, as noted in Figure~\ref{fig:tee_sgx}. This enclave delineates a trusted world where code and data are isolated, maintaining their integrity and confidentiality from the rest of the system. Within this trusted environment, both the application code and data are preserved in an encrypted state and are decrypted solely within the CPU, which guarantees confidentiality. Furthermore, the enclave ensures robust security against external interference from both other applications and system software, irrespective of their privilege levels, whether they operate in the user mode (ring 3) or the kernel mode (ring 0).

\begin {table*}[b]
\centering
\resizebox{.95\linewidth}{!}{
\begin{tabular}{|c|c|c|c|}
\hline
                                     & \textbf{Intel SGX}           & \textbf{AMD SEV}                & \textbf{ARM TrustZone} \\ \hline \hline
\textbf{Processor Architecture}                & x86-64                       & x86-64                          & ARM                    \\ \hline
\textbf{Secure Storage}              & Yes                            & No                               & No                      \\ \hline
\textbf{Remote Attestation}          & Yes                            & Yes                               & No                      \\ \hline
\textbf{Memory Isolation}            & Yes                            & Yes                               & Yes                      \\ \hline
\textbf{Memory Size Limit}           & Up to 128 MB EPC             & Up to available RAM             & 3--5 MB                         \\ \hline
\textbf{Trusted I/O}                 & No                            & Yes                               & Yes                      \\ \hline
\textbf{Operation Level}                        & Ring 3                            & Ring 0                               & Ring -2                     \\ \hline
\textbf{Compatibility}               & Windows                      & Linux-based VMs and hypervisors & Android, Linux         \\ \hline
\textbf{SDK}                         & Provided                     & Not required                    & Provided               \\ \hline
\textbf{Memory Integrity Protection} & Yes                            & No                               & No                      \\ \hline
\textbf{Multithreading}                & Yes  & Yes             & No       \\ \hline

\textbf{Applications}                & Simple and security-sensitive & Complex and legacy              & Lightweight       \\ \hline
\end{tabular}
}
\caption{Comparing properties of the TEE hardware technologies: Intel SGX, AMD SEV, and ARM TrustZone}\label{tbl:hardware}
\end {table*}

SGX creates a limited size of the encrypted memory region, referred to as \textit{Enclave Page Cache (EPC)}, where all the enclaves are created. Based on the hardware access control mechanism, any unauthorized access to the enclave memory is deemed as a page-fault. SGX allows the code inside the enclave to directly access the memory outside EPC, however, such memory accesses are controlled by the operating system (OS) memory management system~\cite{weiser2017sgxio}. Enclaves are also unable to access other enclaves' contents. SGX supports multi-threading within enclaves to accelerate the parallel execution of trusted applications. 

Intel SGX provides \textit{remote attestation} policy that evaluates the enclave identity, integrity of the code inside of it, and guarantees the authenticity of the Intel processor. Remote Attestation serves as a verification
protocol for the service provider to evaluate the health of
enclave(s) created at a remote location~\cite{mofrad2018comparison, van2018foreshadow, chen2019sgxpectre}. SGX also offers \textit{enclave sealing} mechanism that encrypts the enclave to be safely stored in an untrusted storage medium, such as a hard drive, for later use. It also helps the enclave while retrieving data and secrets from the sealed file without performing a new remote attestation~\cite{van2018foreshadow,costan2016intel}.  

While Intel SGX fulfills most of the objectives set forth by the GlobalPlatform TEE, such as secure storage and isolated
execution, it does not provide a native trusted User Interface (UI) or network communication directly from the enclaves. Specifically, the Intel SGX technology focuses on the processor and memory communication aspects. However, SGX fails to provide any feature to facilitate secure communication with the I/O devices. Hence, it is necessary to integrate SGX with other solutions to enable secure communications, for instance, hypervisor-based trusted path architectures with respect to I/O devices~\cite{weiser2017sgxio,liang2019establishing}.

\begin{figure} [!htbp]
\centering
\includegraphics[width=.4\linewidth]{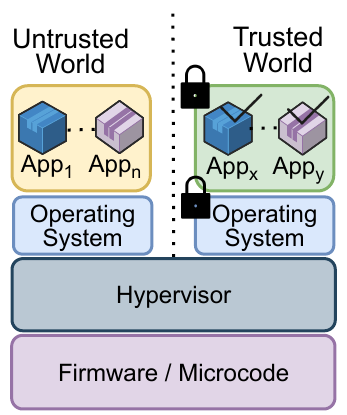}
\caption{High-level architecture of ARM TrustZone. TrustZone provides secure world that facilitates isolated execution. It has access to all hardware resources and can interact with Untrusted/Normal world without tempering privacy.}
\label{fig:arm_archi}
\end{figure}

\subsubsection{ARM TrustZone}
ARM TrustZone is another secure space-based TEE hardware architecture that extends the security aspect to the entire system design, allowing any part of the system to be protected. TrustZone technology provides an infrastructure that
allows chip designers to choose a range of components that can assist with specific functions inside of a
secure environment.

ARM TrustZone architecture can be isolated in two logical states: a secure/trusted world and a normal (\ie insecure) world~\ref{fig:arm_archi}. 
The mechanism that controls information flow between the two states is
called \emph{monitor}. Communication with the secure world occurs from the insecure world through the Secure Monitor Call (SMC) instruction. When an SMC instruction is invoked from the normal world, the CPU core performs a context switch to the secure world and freezes its normal-world execution. All
other CPU cores of a multi-core system can remain in normal-world mode uninterruptedly.
To facilitate secure world processing, TrustZone can separate physical memory into two partitions, with one partition being exclusively accessible by the secure world. 
This isolation is enforced by the memory controller providing access control for memory regions
based on the current state. 
While the normal world cannot access memory allocated to the secure world, the
secure world can access normal-world memory. Therefore, when an application executes in a secure world, TrustZone can
isolate parts of the memory for its use via preventing accessing these locations by the applications executing in the real world~\cite{valadares2021trusted}.

ARM technology is predominantly used in single-purpose systems, such as IoT and edge computing, where the chip is specific to the target market (\eg in smartphones, smartglasses, \etc). Hence, an ARM-based device only has one TrustZone with the processor. This is unlike Intel SGX that take advantage of multiple enclaves.

\subsubsection{Memory Encryption-based TEE: AMD SEV}
AMD  Secure Encrypted Virtualization (SEV) technology is considered to be the most recent hardware assisted TEE that encrypts and protects system memory. The technology has brought  \emph{AES-128} bit encryption engine inside the System on Chip (SoC) that encrypts and decrypts the data upon leaving or entering the SoC.

\begin{figure} [!htbp]
\centering
\includegraphics[width=.4\linewidth]{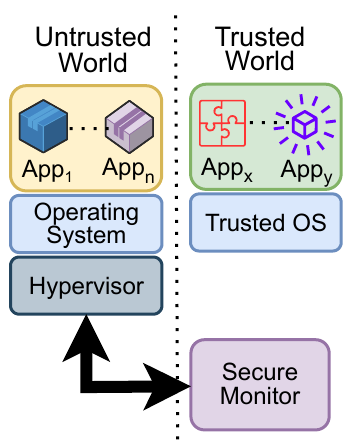}
\caption{High-level architecture of AMD SEV. SEV leverages memory encryption to isolate virtualized environments (\eg containers and VMs) for confidential processing from the underlying platforms.}
\label{fig:sev_archi}
\end{figure}

As shown in Figure~\ref{fig:sev_archi}, SEV isolates virtualized environments (\eg containers and VMs) from the underlying platforms (\eg hypervisor) through memory encryption. Although hypervisors are commonly used as trusted components in the virtualization security model, they cannot guarantee the security of confidential workloads.
For instance, to preserve data/workload confidentiality in the cloud, users need to secure their VM-based workloads from the cloud provider (administrator). This leads to the necessity of hardware level VM isolation which is what SEV can fulfill. 
SEV allows a single VM to be assigned a unique AES encryption key to encrypt the data they use. Consequently, even if the hypervisor tries to read the memory inside the guest OS, it can only fetch the encrypted bytes.
AES encryption provides increased confidentiality protection of memory. An attacker without proper knowledge of the encryption key cannot decipher VM data. Note that SEVs memory encryption keys are generated from a hardware random number generator and is stored in dedicated hardware registers that cannot be directly read by systems. In addition, the hardware is designed in such a way that the same plain-text is encrypted differently in different memory locations~\cite{AmdSev}.

AMD SEV does not require any modifications
to the user application software and the memory encryption is transparent to the application executing in the SEV-protected VM. SEV uses the AMD Memory Encryption Engine which is capable of working with different encryption keys for encrypting and decrypting different VM memory spaces on the same platform.

In SEV, a unique encryption key is associated with each guest VM. Upon code and data arrival to the SoC, SEV tags all of the code and data associated with the guest VM in the cache and limits the access only to the tag’s owner VM. Upon data leaving the SoC, the VM encryption key is identified by the tag value and data is encrypted with the VM key. Moreover, initializing a SEV-protected VM requires direct interaction with the AMD secure processor. The AMD secure processor provides a set of APIs for provisioning and managing the platform in the cloud. The hypervisor’s SEV driver can invoke these APIs.

In the AMD SEV architecture, the guest owner manages the guest secrets and generates the policies for VM migration or debugging. The Diffie-Hellman key exchange protocol 
is used between the guest owner and the AMD secure processor to open a secure channel between the guest owner and the AMD secure processor. The guest owner is enabled to authenticate the secure
processor and exchange information to set up the protected VM. Also, the SEV architecture defines the shared page (unencrypted) and the private page (encrypted) that can be set for each protected guest VM. The C-bit is set to identify the private pages by the guest
OS. 

\subsubsection{Other TEE Implementations}
While Intel SGX, AMD SEV, and ARM TrustZone have gained significant attention in the confidential computing realm, several other TEE implementations also hold considerable promise to realize confidential computing. Here are some notable mentions:

\textit{RISC-V:} RISC-V \cite{riscfive} is an open standard instruction set architecture that is growing in popularity for its flexibility and extensibility. Several initiatives (\eg \cite{nashimoto2022bypassing,nashimoto2022poctee,hex5}) are underway to develop TEE implementations based on the RISC-V architecture, aiming to provide hardware-level security guarantees while preserving the open and customizable nature of RISC-V.

\textit{Open Enclave SDK:} Microsoft's Open Enclave SDK~\cite{openenclave} is a library that allows developers to create applications that leverage TEEs across different hardware platforms. It abstracts the specifics of the underlying TEE technology, making it easier for developers to create confidential computing solutions without the need to understand the complexities of each hardware solution.

\textit{Keystone:} Keystone~\cite{keystone} is an open-source project that aims at building a secure and customizable enclave based on RISC-V the architecture. It strives to provide an open, flexible, and extensible TEE that is verifiable by the community.

\subsubsection{Advancements in the TEE Adoption}
The integration and adoption of Trusted Execution Environments (TEEs) have made significant strides over recent years, spanning various applications and addressing multiple challenges. The research works undertaken by Zhang \etal \cite{zhang2016sok} and Ning \etal \cite{ning2017position} survey existing TEEs, highlighting their utility and potential roadblocks. 
Furthermore, the evolution of TEEs has been driven by the need for robust security measures in modern defense systems. As exemplified by MemSentry \cite{koning2017no}, TEEs are being used to leverage hardware features to improve overall system security. The implementation of TEEs in this regard has revolutionized defense system security, making TEEs an integral part of contemporary defense technologies.

Adoption of TEEs in data analytics and debugging processes is also gaining momentum. This is illustrated by the work of Schuster \etal \cite{schuster2015vc3}, where Intel SGX was utilized to secure data analytics. Similarly, the work of Ning \etal \cite{ning2017ninja} leverages the ARM TrustZone technology to enhance transparency in debugging and tracing processes.

In light of these advancements, the practical adoption of TEEs has extended beyond theoretical applications, becoming a crucial tool in mobile, wearable systems, data analytics, defense technologies, and debugging processes. Continued work in this area is expected to drive further advancements, cementing the role of TEEs in ensuring secure and efficient solutions in distributed systems and, particularly, within the context of IoT and edge-to-cloud continuum.

\subsection{Hardware Modules for Confidential Computing}
\label{hwmodules}
\subsubsection{Trusted Platform Module (TPM)}
TPM is a dedicated secure co-processor chip that is designed to carry out cryptographic operations (\ie digital rights managements) via ensuring safer computing across multiple environments. TPMs are placed on the motherboards to provide trusted computing capabilities to the system.
The chip includes a set of security mechanisms to construct it as temper-resistant, so that malicious programs unable to tamper the security codes of the TPM. Each TPM contains an RSA key pair, called the the Endorsement Key (EK) that is generated during the manufacturing process of the TPM chip. Each chip owns a unique and identifiable EK that is managed within the chip and is inaccessible by any software. When a system user/administrator takes ownership of the system, The Storage Root Key (SRK) is created based on EK and the owner-specified password. 

Besides EK, TPMs generate another key pair, named Attestation Identity Keys (AIK), that protect the device against unauthorized firmware or software manipulation via hashing critical sections of them before execution.
When the system attempts to connect to the network, the hashes are sent to an authentication server for verification purpose. If any hashed components is compromised since the last execution, the network access authentication fails. 

The key advantages of adopting TPM technology for confidential computing are twofold:
\begin{enumerate}
    \item Generating, storing, and controlling the use of cryptographic keys.
    \item Creating an unforgeable hash key summary of the hardware and software configuration, digital right management, and software licensing. Via examining the hash key, a third party can verify the integrity of the software.
\end{enumerate}

The initial version of TPM 1.2 was released in 2005 and it has been updated to TPM 2.0 in 2019 with a wider range of security features. The advancements of TPM 2.0 over the previous version are as follows:
\begin{itemize}
    \item \textit{Making use of newer algorithms}: TPM 1.2 leverages SHA-1 hashing algorithm that raises security concerns. Hence, SHA-256 algorithm was adopted and TPM 2.0 now supports a variety of newer algorithms that improve the performance of drive signing (\ie signing device drivers) and key generation.  
    \item \textit{Supporting more data types}: TPM 1.2 only supports unstructured data in NVRAM, whereas, TPM 2.0 supports unstructured data, Counter, Bitmap, Extend, and PIN pass/fail.
    \item \textit{Supporting more hierarchical structures}: TPM 1.2 only has a storage hierarchy, whereas, TPM 2.0 supports three hierarchical structures that are: (A) Platform Hierarchy (PH) \cite{phier} that represents the root of trust for the platform and is typically controlled by the platform manufacturer. The PH is responsible for managing platform-specific operations, such as initializing the TPM and controlling its critical functions. (B) Storage Hierarchy (SH) that is in charge of managing keys and authorizations related to the storage and retrieval of sensitive data within the TPM. (C) Endorsement Hierarchy (EH) which is mainly used to establish the identity and authenticity of the TPM via creating certificates, signing them using the EK, and participate in the attestation protocols to prove the platform's trustworthiness.
    
\end{itemize}

\paragraph{Various Implementations of TPM:}
\begin{itemize}

 \item \emph{A discrete TPM} is implemented as a separate function or feature chip, with the required computing resources that are contained within the discrete physical chip package. A discrete TPM has full control of dedicated internal resources such as volatile and nonvolatile memory, cryptographic logic, \etc and it can only access and use these resources. Hence, they are considered the most secure type of TPM. Intel Trusted eXecution Technology (TXT) \cite{inteltxt} and Platform Trust Technology (PTT) \cite{intptt} are some implementations of discrete TPM 2.0.
     
  \item \emph{Integrated TPMs} are implemented as a dedicated hardware that are integrated into embedded into the hardware of a computing device (\eg laptop or server motherboard). However, they are logically separate from other components of the system. 
  \item \textit{A firmware-based TPM (fTPM)} \cite{ftpm} is a software implementation of a TPM that resides in the firmware of a computing device. It emulates the behavior and features of a hardware-based TPM using firmware code. It can also operate using the resources of a multi-feature compute device such as SoC CPU in the context of a TEE. An fTPM does not have its own dedicated storage, thus, it relies on the operating system and relevant platform services to get the storage access right. 
    
    \item \textit{Hypervisor TPMs} \cite{vtpm} are virtual TPMs (vTPMs) provided by hypervisors, hence, they are dependent on the hypervisors. The vTPMs run in an isolated execution environment that is hidden from the other software applications executing inside VMs. The aim of such special execution scheme is to secure their code from the software in the VMs. The vTPMs offer a security level comparable to fTPMs. Google Cloud Platform has implemented and utilizes vTPMs in its offerings~\cite{techtpm}.
    \end{itemize}

\subsubsection{Secure Element (SE): Hardware Module for Confidential Computing}
A Secure Element is a tamper-resistant platform that is capable of hosting programs (\eg code and script) and confidential data, such
as cryptographic keys, securely according to the security processes set forth by its owner.
The concept of a secure element came to light in the mid of 1970s in the form of a smart card which was based on a one-chip. A secure micro-controller running a secure ultra-light operating system, which was restricted to execute only one application for a long time~\cite{shepherd2016secure}.

\begin{figure} [!htbp]
\centering
\includegraphics[width=.72\linewidth]{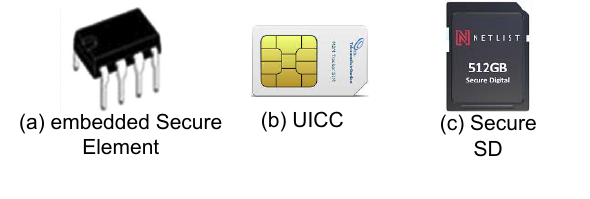}
\caption{Modern forms of Secure Elements (SEs). Part (a), Embedded Secure ELement (eSE), is used in the smartphone hardware. Part (b), UICC, is for secure identification and authentication of users in cellular (mobile phone) networks. Part (c), Secure SD card, provides additional security layer via enabling the secure data storage and processing within the card.  }
\label{fig:secard}
\end{figure}

In contrast to TPMs, SEs are able to execute secure code and are not restricted to perform only cryptographic operations. Smart cards were the only type of SE used with different types of connectivity, such as USB dongles and contactless smart cards over a long time period. Recent invention of Near Field Communication (NFC) \cite{nfc} introduces three new forms of SE that are shown in Figure~\ref{fig:secard}. These SEs are as follows: 
\begin{itemize}
\item Embedded SE (eSE) \cite{selement} is a smart card micro-controller integrated into the NFC chip or directly used in the hardware of the mobile phone. It provides a secure environment for the management, storage, and processing of confidential and cryptographic data. It is typically integrated into the Near Field Communication (NFC) chip of mobile devices or directly into their hardware, forming an essential component of confidential computing. The eSE protects sensitive data such as encryption keys, digital certificates, and user credentials. In applications such as mobile payments, access control, or authorization, the eSE performs secure transactions by using the stored data to authenticate interactions, thereby offering a secure and confidential computing environment.
    \item Universal Integrated Circuit Card (UICC) \cite{uicc}, commonly referred to as a SIM card, contains mobile subscriber identity data to facilitate secure user identification and authentication on mobile devices. It establishes the security groundwork for mobile devices by protecting user identity information, network authorization metadata, and personal security keys. The UICC enables secure network connection and confidential communication, effectively preserving data integrity and privacy.
    \item Secure SD card \cite{aesmem} is another type of smart card micro-controller that enables features such as secure data storage, hardware encryption, and tamper-resistant processing. These cards operate by utilizing hardware encryption algorithms to secure data stored on them. This ensures the confidentiality of the data, rendering it unreadable without the corresponding encryption key, even when the card is relocated to another device. As a result, Secure SD cards offer a robust solution for secure and portable storage of sensitive data in the realm of confidential computing.
\end{itemize}
 


\section{Confidential Computing Middleware across Edge-to-Cloud}
\label{sec:middleware}
Middleware in any form of distributed system, including edge-to-cloud, is an abstraction layer that lies between the OS of each machine and the applications executing across the distributed system. The aim of the middleware is to hide system complexity and enable seamless sharing of resources between users and applications. A middleware-based distributed system is susceptible to attack, as anyone with the root access to the worker machines is able to inspect, modify, kill, or modify code or data executing on that machine. Hence, to establish confidential computing across edge-cloud, enabling trusted computing on the middleware is a necessity. 
Confidential Computing addresses this sort of attacks via maintaining the confidentiality during the execution of a software irrespective to the privilege levels of any individual who has access to the system. 
In this Section, we will discuss the scopes in the middleware where confidential computing is performed. 

\subsection{Confidential Computing: Edge-to-Cloud Perspective}
\label{subsec:trust edge cloud}
The cloud computing paradigm offers large scale managed sharing and interpolation among the dispersedly controlled resources. 
However, data security and privacy concerns are the crucial factors that have made many organizations reluctant to use cloud \cite{woodworth2019s3bd}.
As example, the cloud infrastructures are often vulnerable to insider threats, such as former employees accesses, manipulating, or destroying a copy of confidential data (including on-site backups).
 Numerous recent data privacy violations~\cite{zobaedbig} in the cloud environments have raised serious data privacy concerns. 
 In one incident~\cite{verizonacci}, more than 14 million Verizon customer accounts information were leaked from their cloud repository. In another incident~\cite{yahooacci}, confidential information of over three billion Yahoo users were leaked. 
Consequently, the cloud beneficiaries (\ie individuals, organizations) with sensitive data are hesitant to fully embrace the cloud paradigm due to the privacy concerns. Users expect that all their processing and the communications in the cloud are trusted. In practice, a cloud provider explains the compliance and service level agreement (SLA) to beneficiaries, however, that could not ``guarantee'' technical enforcement or transparency such that the beneficiaries can safely run their sensitive workload on the cloud.

A trusted cloud infrastructure is expected to provide increased reliability, technical enforcement, and security assurances. Confidential cloud computing (CCC) meets these expectations via allowing the beneficiaries to specifically define the required hardware and software that have access to their workloads (\ie data and applications). Adopting confidential computing provider users with the full control over their workloads, software, and hardware systems. It prevents cloud-hosting infrastructures (\eg hypervisors~\cite{davoodpaper}) access to their sensitive data. Establishing CCC depends on the cloud computing paradigm, namely \textit{serverful} (elaborated in the next section) vs \textit{serverless} \cite{denninnart2021harnessing}, and the middleware used to offer the services in each paradigm. Moreover, it depends to the type of isolation (virtualization) that is offered in each paradigm \cite{davoodpaper1}.

Edge computing has emerged to fill the gap between client machines and remote cloud datacenters. Being deployed near to the user, edge systems have are able to handle data-driven and/or compute-intensive applications, such as augmented reality, video analytics, and ML. As we move from the cloud to user premises, communication latency decreases, but trustworthiness increases~\cite{zobaed2023ai}. 
As the edge nodes are close to the end user’s
premises, they are often deemed trustworthy and data privacy is ensured to match with the computing requirements.  A large body of research (\eg~\cite{grassi2017parkmaster,  wu2017edge,  yi2017lavea,  chen2018industrial,  zhang2016demo, qi2017vehicle}) have been conducted in the usage scenarios and performance of edge computing. However, the privacy and security of the edge computing are not much dealt with.  

\subsection{Confidential Computing in the Serverful Cloud Paradigm}
The serverful computing paradigm refers to the conventional form of resource provisioning in the cloud that is based on Bare Metal (BM) servers, VMs, or containers listening for requests on port 80. 

\subsubsection{Bare Metal (BM)} BM resource provisioning describes an environment where physical dedicated servers are provisioned to customers. The important point is that BM servers do not use any form of virtualization and hypervisor~\cite{zhang2020high}. Consequently, BM users have full control over the allocated servers, including its processing, storage, and networking subsystems~\cite{linkbarem}. This is not the case for virtualized (multi-tenant) servers running on a shared hardware. As such, BM users have the freedom of configuring any trusted OS environments and applications as well as installing hypervisors to create their own VMs to satisfy their requirements.

Deploying a trusted application on the BM server requires attention to the security of the underlying platform, including using a trusted hardware, patched OS, and controlling access permissions. At the hardware level, making use of a trusted execution environment (TEE) is a must and, for that purpose, cloud providers often rely on Intel SGX and AMD SEV to offer confidential computing in the execution of trusted applications~\cite{anjunabm}. Popular cloud service providers such as IBM, Alibaba, and Platform9 clouds have configured Intel SGX and offer it in their BM deployments. In~\cite{iBMbmprov}, IBM provides complete documentation of provisioning a fully SGX-based BM server. 

\begin{figure} [h]
\centering
\includegraphics[width=.5\linewidth]{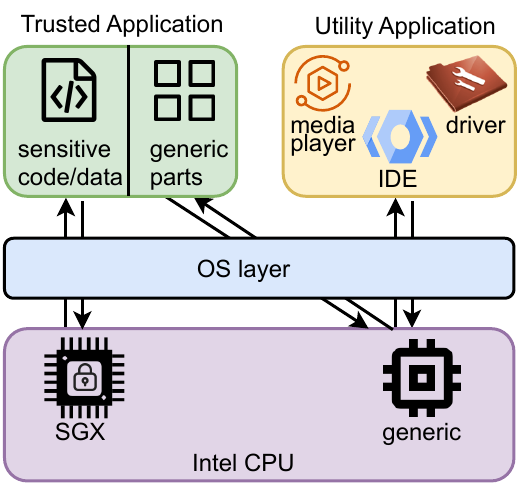}
\caption{Trusted Application setup on top of the Intel SGX-based bare metal (BM) server}
\label{fig:sgxbm}
\end{figure}

Figure~\ref{fig:sgxbm} shows a high-level diagram representing the Intel SGX application setup in a BM server of the IBM cloud. According to this figure, a trusted application composed of generic and sensitive code/data, respectively, can make use of the generic and SGX cores of the Intel CPU.
Such implementation facilitates confidential computing to the extent that even an attacker with root privilege to the BM instance cannot access or tamper the code, data, or the returned outputs. In addition, even the cloud provider cannot access or tamper with the code/data, despite their direct access to the hardware as well as root access to the host OS~\cite{anjunabm}. 

\subsubsection{Confidential Virtual Machines (VMs)}
A VM provides an isolated and exclusive execution environment by running its own OS and functions separately from the other VMs, even when they run on the same physical host machine. VMs serve many use cases as they can be deployed across on-premises and cloud environments. Particularly, public cloud providers are using VMs to provide cost-efficient and flexible (with root access) virtual resources to multiple users at the same time (a.k.a. multi-tenancy). In this case, VMs are hosted on remote servers that are not under the control of the VMs' owners (users). Hence, the trustworthiness becomes a major challenge for cloud VMs.

For the purpose of confidential computing, \textit{confidential VMs} have been proposed to run alongside of other standard VMs atop hypervisor. By definition, VMs inherently maintain a high degree of isolation from each other. In addition to this inherent isolation, a confidential VM is protected by hardware-based encryption keys that prevent a malicious VM manager to breach its confidentiality.
Confidential VMs track the system record in the background for attestation purposes and to verify the system security. Although it is essential to deploy confidential VMs on top of TEEs  for fast and easy adoption, it raises some challenges. For example, the VM administrator has full read/write control over the it, which is too coarse in many cases. Another concern is that the VM TCB is large. This because the VM image is far more than just a kernel and an application; it includes a large number of system services. In the worst case scenario, this configuration still proves to be more secure than running the software on-premises or on the existing cloud infrastructure. 

\noindent\textbf{{Confidential VM via AMD SEV.}} AMD SEV is one of the most common TEE technologies that is used as the hardware to deploy confidential VM. To provide runtime protection, AMD's memory encryption engine encrypts the memory contents of the SEV-enabled confidential VMs using AES-128 encryption algorithm. An integrity-protected firmware deployed on top of a dedicated co-processor called Platform Security Processor (PSP) is responsible to generate the required encryption keys. The co-processor has its own memory and nonvolatile storage while having access to the system memory of the main processor. Moreover, the integrity-protected firmware provides APIs that can be used by the host hypervisor for encryption key managements on behalf of all SEV-enabled confidential VMs running on that system. The APIs also handle secure data transfer between the host hypervisor and the virtual memory of a guest confidential VM. It is noteworthy that AMD-SEV does not exploit any software, therefore, developers do not require any AMD APIs or libraries  to make their applications compatible.

Nevertheless, because cloud are deemed as untrusted providers who can modify a VM while it is being deployed, the runtime protection does not ``guarantee'' the confidentiality of workloads executed on an SEV-enabled VM. As such, remote attestation feature can be used to establish trustworthiness. This feature can be used to verify an authentic SEV-enabled AMD platform configuration for the VMs on the cloud.
At present, AMD-SEV is only available in AMD's EPYC series processors that are intended only for servers.

\noindent\textbf{{Confidential VM via Intel SGX.}} 
Confidential VMs are also configured on a special implementation of Intel SGX, called \textit{vSGX}, that enables VMs to use Intel SGX technology if it is available on the hardware.
To use vSGX, the bare metal hypervisor host (\eg ESXi) needs to be installed on an SGX-enabled CPU and vSGX functionality is enabled in the BIOS of the host machine. 
Despite AMD SEV, it is observed that Intel SGX-enabled VMs require considerable amount of effort to make the code compatible.
Unlike AMD SEV, however, Intel SGX requires the programmers to use the Intel-SDK to specify which parts of the application will be executed on the trusted or untrusted subsystems. 

SGX's isolated memory regions are ideal for small-TCB (Trusted Computing Base) services, even though using them to run confidential VMs is challenging. This is mainly because of two reasons: (i) VMs need large TCBs, because a VM image is significantly larger than just a kernel and an application. (ii) Lack of support for multiple address spaces and privileged and unprivileged mode separation. Provided these facts, \textit{we can conclude that AMD SEV is more applicable for confidential VMs than SGX}.






\subsubsection{Confidential Containers}
Container is a software package that contains all of the required components (such as file systems, libraries, environment variables, \etc) to execute applications in any environment without having side-effects on other applications on the same machine. As containers can deploy and execute applications in isolation that access a shared OS kernel, they are comparable to VMs. Therefore, containers are used as a replacement (or complement) of VMs where the allocation of hardware resources is carried out through containers.

Containers are lightweight and orchestrated to virtualize single applications. They create isolation boundaries at the application level, not at the server level. This isolation means that if anything goes wrong  (\ie a process consuming too many resources, unexpected exceptions) in a particular container, it will only affect that container, not the entire VM or server.  It also eliminates compatibility issues between containerized applications on the same operating system.

The containers that execute confidential workloads need to be isolated, secured (encrypted), and inaccessible so that it can be protected from misuse. That is why, the idea of \textit{confidential container} has emerged to conceptualize running a container within a hardware TEE platform, thereby, ensuring protection against vulnerabilities at the guest OS, hypervisor, and host OS levels. In this case, because of the TEE involvement, a confidential container provides a set of features alongside of an existing container deployment to make it secured, encrypted, and more isolated, thereby, achieving a higher data security. Confidential containers can mitigate the limitation of confidential VMs via offering finer degree of control and ensuring faster and secure execution of containerized applications. It also allows running standard container images with no modification or recompilation in code within a TEE setup.

\begin{figure} [h]
\centering
\includegraphics[width=.7\linewidth]{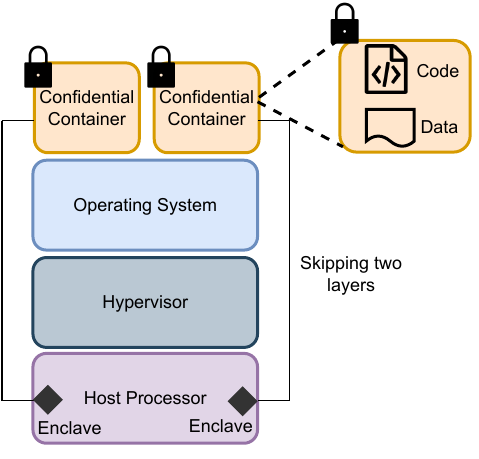}
\caption{Layered view of the confidential container deployment in Microsoft Azure. Azure leverages SGX to offer direct execution to the host processor via bypassing the guest OS, host OS, or hypervisor from the trust boundary. }
\label{fig:secure_enc}
\end{figure}

Kubernetes \cite{kubernetes}, a widely adopted portable and open-source platform, used for managing (a.k.a. orchestrating) the containerized workloads and services, thereby, mitigating the burden to rely on and trust the system administrators to securely launch containers. More specifically, upon launching a confidential container on a Kubernetes node that is capable of performing confidential computing, it creates a process-isolated and \textit{sandboxed enclave}. By definition, \textit{Enclaves are secured portions of the TEE's processor and memory}. The memory allocated by the enclave is process-specific, allowing for isolation across containers and protection for each container. Through container encryption, signing, and attestation, enclaves ensure code integrity and prevent malicious attacks that might attempt to tamper with the code inside the container.

Figure~\ref{fig:secure_enc} illustrates deployment architecture of confidential containers in Microsoft Azure Kubernetes service. Azure leverages the Intel SGX processor, which allows user-level code from containers to allocate private memory regions to execute the code. The discrete execution model per container per node allows the hardware to run applications directly on the host processor and encode a dedicated block of memory within a single container.
To launch existing docker containers, applications on confidential compute nodes require Intel SGX wrapper software to help containers execute with the special CPU instruction set. SGX creates a direct execution to the host processor via bypassing the guest OS, host OS, or hypervisor from the trust boundary. This step reduces the overall attack surface and vulnerabilities, while enabling process-level isolation within a single node.

\subsection{Confidential Computing in the Serverless Cloud Paradigm}
Serverless computing is an execution model in which a cloud provider dynamically allocates---and charges the user for---only the compute resources and storage needed to execute a particular piece of code (a.k.a. function). Serverless computing paradigm enables developers to concentrate on the business logic by writing fine-grained and standalone functions with minimal overhead of the deployment, management, and scalability. 
Naturally, there are servers involved in the back-end, however, their provisioning and maintenance are entirely handled by the provider and is transparent from the user's perspective. The paradigm aims at mitigating the job of cloud programmers \cite{chavitsurvey} and is considered as the next generation of cloud computing system. Consequently, hyperscaler clouds such as Amazon, Microsoft, and Google have introduced AWS Lambda~\cite{serverexa}, Azure Functions~\cite{azure}, and Google Cloud Functions,~\cite{serverexg}, respectively.  
According to the recent research~\cite {shahrad2020serverless,li2021confidential}, around $54\%$ serverless applications contain only one function whereas $50\%$ of them have less than one second of executing latency.  
Serverless paradigm can be defined as the aggregation of Function-as-a Service (FaaS) and Backend-as-a Service (BaaS) \cite{jindal2021estimating}. While FaaS focuses on the front-end development of functions, BaaS focuses on the transparent and isolated execution of the functions \cite{chavitsurvey}.

Alongside the regular executions, serverless computing is used to execute privacy-preserving workloads such as authentication, personalized chatbot, biometrics (\eg face/fingerprint recognition) processing. To protect the user privacy in the serverless paradigm from different security threats, including malicious cloud software and suspicious cloud insider, it is essential to enable confidential computing within this paradigm. Secure enclaves, offered by Intel SGX, are widely used as a trusted hardware component to offer a fully isolated execution environment for privacy-preserving serverless applications. However, the existing CC architectures are not well-suited for confidential processing, as they cause performance degradation up to $422.6\%$~\cite{li2021confidential}. The majority of this overhead is causes by the  enclave initialization: hardware enclave creation and attestation measurement
generation during the startups and secret data transfer among the functions during their executions. Li \etal show that secret data transfer between functions can take up to 34.7\% of the execution time~\cite{li2021confidential}. Although recent studies~\cite{trach2019clemmys,weisse2017regaining,orenbach2017eleos} propose different software-based optimization techniques, the end-to-end latency to invoke a function is nowhere close to the satisfactory range. Li \etal conclude that the design technology of the current SGX is the reason for the high latency overhead. According to their paper, the existing SGX hardware architecture is not featured with memory sharing option across the enclave instances~\cite{li2021confidential}. 

Although SGX enclaves ensure security, such excessive start-up latency (cold-start) is ill-suited for today's confidential serverless computing. To improvise enclave abstraction, Li \etal extend the existing SGX implementation with \emph{region-wise plugin enclaves} that can be immutably mapped into different existing isolated enclaves to reuse attested common states across the functions~\cite{li2021confidential}. They instrument the dynamic resizing policy of SGX to implement a hardware-enforced copy-on-write mechanism for maintaining consistency between the content and measurement of a plugin enclave. They show that the plugin enclave-based serverless can scale up to $10\times$ enclave instances than existing SGX hardware
while reducing $\approx 96\%$ overhead of function startup latency. 

In~\cite{feng2021scalable}, Feng \etal enable a fast enclave creation without compromising the security guarantees. They propose the notion of Guarded Page Table (GPT) to enable memory isolation with page-level granularity and Mountable Merkle Tree (MMT). Together, these concepts can achieve on-demand abstraction and integrity protection. Upon applying these two concepts, they propose the notion of \emph{shadow enclave} that supports fast enclave creation without compromising security guarantees. Their experimental evaluations indicate that the shadow enclave system is capable to scale to thousands of concurrent secure enclaves with proper resource utilization and reduce the start-up latency by three orders of magnitude~\cite{feng2021scalable}.


\subsection{Confidential Computing on the IoT (Device) Tier}
Unarguably, establishing confidential computing across edge-to-cloud entails enabling it on each tier contributing to the continuum. Despite differences in the scale of deployed resources on the edge, fog, and cloud, the confidential computing solutions we discussed in the previous two parts (\ie serverless and serverful paradigms) overlap across these tiers. This is because: (a) all these tiers predominantly operate based on some standard middleware solutions, such as Kubernetes~\cite{kubernetes} and OpenStack~\cite{openstack}; (b) similar isolation techniques (\eg VMs, containers, \etc) are employed across these tiers; and (c) servers of these tiers are configured with widely-used operating systems, which is often some flavor of Linux.

The IoT device tier, however, does not share the aforementioned three similarities with the other tiers. In practice, IoT devices are categorized as embedded systems \cite{ayoade2019secure} that often have their own custom firmware. Moreover, they are highly distributed and accessible that makes them more vulnerable. These qualities imply different type of threats and confidentiality solutions for the IoT tier. Accordingly, in this section, we concentrate on the traits of IoT devices and challenges and solutions of enabling confidential computing on them.

Ultimately, the comprehensive \textit{IoT standardization} will be pivotal in fully realizing the potential of confidential computing at the IoT tier \cite{kugler2023standards}. Currently, there are some fragmented efforts for IoT standardization (\eg ISO/IEC 30141 \cite{di2018internet} and  TS 103645 \cite{cirne2022iot}) that prescribe issues like using a common vocabulary and reusable designs for IoT devices. However, there is yet to be a consensus on these standards and this discrepancy is the Achilles' heel for the \cc\; across the entire continuum. That is why, organizations are actively seeking for alternative security mitigation strategies.

First and foremost, organizations should consider zero-trust access (\textit{ZTA}) to verify users' and devices before every application session. This assures that the users and devices meet the organization's policy to access that application, therefore, dramatically mitigates IoT-level risks.

\textit{Micro-segmentation}~\cite{wasicek2020future} is deemed as another strategy with a significant potential to mitigate vulnerabilities caused by the distributed nature of the IoT tier. The concept of micro-segmentation strategically fragments the expansive IoT network into isolated `micro-segments'. Each of these micro-segments serves as a secured endpoint and is strictly maintained by well-defined security policies. This structure allows for grouping of the IoTs based on their distinct roles within the network and the sensitivity level of the data they manipulate. This is achieved via localizing the potential threats to individual segments, thereby, mitigating the risk of widespread network compromise and enhancement of the confidential computing across the continuum. 

The dynamic nature of IoT networks, with devices of varied security capabilities continuously joining and leaving, presents another level of complexity. However, with security policy identification for each micro-segment, localized threat detection and response mechanisms, and efficient application partitioning across heterogeneous TEEs, this complexity can be handled. 

It is noteworthy that, in addition to these mitigation strategies, basic solutions (such as connection encryption) can still help and are necessary to mitigate the risks of the IoT tier.


\section{Trusted Application Development across Edge-to-Cloud}
\label{sec:application}
\subsection{Confidential Machine Learning (ConfML)}
Machine Learning (ML) can be broadly defined as methods to automate pattern learning from massive datasets to the extent that data analysts no longer need to manually identify the hidden characteristics and correlations of the data. Although ML-based applications are getting wide adoption and are envisaged to revolutionize our world, their algorithms do not ensure confidentiality of the sensitive data/workloads. 

To address the data exposure risk and lack of confidentiality, enabling \cc for the ML workloads is crucial. Recent advancements in TEEs in both high-end and low-end mobile devices make them a prime contender for achieving confidentiality and integrity in ML. Towards that, Confidential Machine Learning, a.k.a. \emph{ConfML} protocol, has been proposed where the data owners must adhere to when from when they share training data with an ML service \cite{graepel2012ml}. This protocol protects the privacy of training data during the training process.

Conventionally, encryption can safeguard the confidentiality of both data at rest and data in transit. Using encryption for ML, however, requires decrypting the data before the start of training which implies data vulnerability until the end of training. ConfML eliminates this vulnerability via protecting the privacy of training data during the training process \cite{graepel2012ml}. The ConfML protocol consists of two phases that bookend the training process: \textit{(i)} The data owner encrypts the training data files with a secret key before submitting them to the ML service. The secret-key is not accessible to the ML service. \textit{(ii)} After acquiring the network-trained-on-scrambled-data from the ML service, the data owner modifies the network to behave as if it were trained on the original (unscrambled) data using the secret-key from step (i). These two procedures ensure that the ML service never has access to the original data, while the data owners obtain the necessary networks.

\begin{figure} [h]
\centering
\includegraphics[width=.8\linewidth]{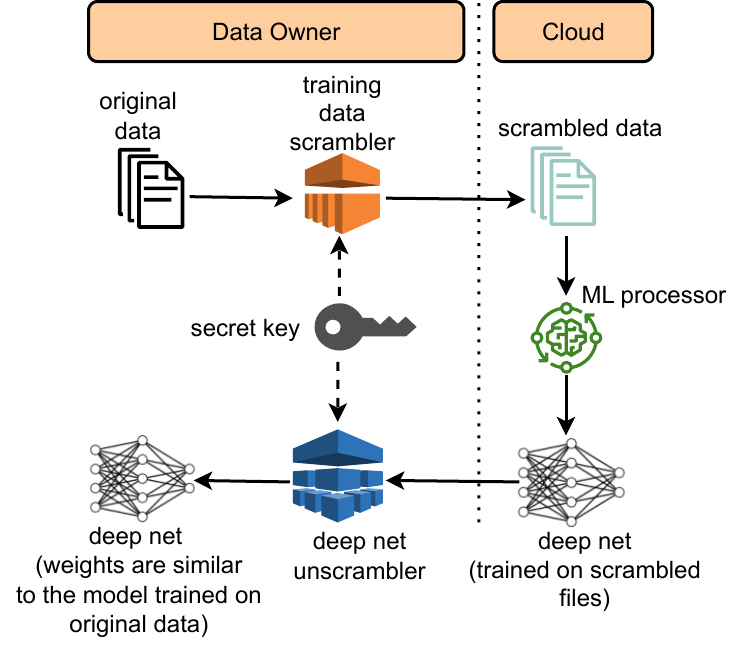}
\caption{Workflow of confidential Machine Learning (ConfML). The workflow consists of two phases of the training process: Data owner and Cloud site. The data owner encrypts the training data with a private key and sends it to the cloud-resided ML service. The trained model is then converted by the owner using this key to function as though trained on original data. }
\label{fig:confml}
\end{figure}

\begin{table*}[]
\resizebox{\linewidth}{!}{
\begin{tabular}{|l|l|l|l|l|l|lll|l|}
\hline
\multirow{2}{*}{\textbf{Prior Studies}} & \multirow{2}{*}{\textbf{Hardware}} & \multirow{2}{*}{\textbf{\begin{tabular}[c]{@{}l@{}}Computing\\ Paradigm\end{tabular}}} & \multirow{2}{*}{\textbf{SDK}} & \multirow{2}{*}{\textbf{DL Library}} & \multirow{2}{*}{\textbf{Training}} & \multicolumn{3}{l|}{\textbf{Protection Objective}} & \multirow{2}{*}{\textbf{Contribution}} \\ \cline{7-9}
 &  &  &  &  &  & \multicolumn{1}{l|}{\begin{tabular}[c]{@{}l@{}}Data Priv-\\ acy\end{tabular}} & \multicolumn{1}{l|}{\begin{tabular}[c]{@{}l@{}}Model Conf-\\ identiality\end{tabular}} & \begin{tabular}[c]{@{}l@{}}Training\\ Integrity\end{tabular} &  \\ \hline
\begin{tabular}[c]{@{}l@{}}Ohrimenko~\etal\\~\cite{ohrimenko2016oblivious}  (2016)\end{tabular} & SGX & Cloud & SGX SDK & Fast CNN & Y & \multicolumn{1}{l|}{Y} & \multicolumn{1}{l|}{N} & N & \begin{tabular}[c]{@{}l@{}}- Data-oblivious ML \\ for classification \\ and clustering\end{tabular} \\ \hline
\begin{tabular}[c]{@{}l@{}}Hunt~\etal~\cite{hunt2018chiron}\\ (2018)\end{tabular} & SGX & Cloud & SGX SDK & Theano & Y & \multicolumn{1}{l|}{Y} & \multicolumn{1}{l|}{Y} & N & \begin{tabular}[c]{@{}l@{}}- ML-as-a-service \\ implementation\\ on multi enclaves\end{tabular} \\ \hline
\begin{tabular}[c]{@{}l@{}}Lee~\etal~\cite{lee2019occlumency}\\ (2019)\end{tabular} & SGX & Edge-Cloud & SGX SDK & Caffe & N & \multicolumn{1}{l|}{Y} & \multicolumn{1}{l|}{Y} & N/A & \begin{tabular}[c]{@{}l@{}}- On demand loading\\ - Channel partitioning\end{tabular} \\ \hline
\begin{tabular}[c]{@{}l@{}}Mo~\etal~\cite{mo2020darknetz}\\ (2020)\end{tabular} & TrustZone & Edge & OP TEE & Darknet & Y & \multicolumn{1}{l|}{N} & \multicolumn{1}{l|}{Y} & N & \begin{tabular}[c]{@{}l@{}}- Privacy measurement\\ - Layer-wise partitioning\end{tabular} \\ \hline
\begin{tabular}[c]{@{}l@{}}Liu~\etal~\cite{liu2021trusted}\\ (2021)\end{tabular} & TrustZone & Cloud & OP-TEE & - & N & \multicolumn{1}{l|}{N} & \multicolumn{1}{l|}{Y} & N/A & \begin{tabular}[c]{@{}l@{}}- Weights \& Feature-\\ map partition\end{tabular} \\ \hline

\begin{tabular}[c]{@{}l@{}}Zhang~\etal~\cite{zhang2021citadel}\\ (2021)\end{tabular} & SGX & Cloud & SCONE & Tensorflow & Y & \multicolumn{1}{l|}{Y} & \multicolumn{1}{l|}{Y} & N & \begin{tabular}[c]{@{}l@{}}- Multi enclaves for\\ training\end{tabular} \\ \hline
\begin{tabular}[c]{@{}l@{}}Mo~\etal~\cite{mo2022ppfl}\\ (2021)\end{tabular} & \begin{tabular}[c]{@{}l@{}}SGX + \\ TrustZone\end{tabular} & Edge-Cloud & OP-TEE & Darknet & Y & \multicolumn{1}{l|}{Y} & \multicolumn{1}{l|}{Y} & Y & \begin{tabular}[c]{@{}l@{}}- Layer-wise training\\ - Privacy measure\end{tabular} \\ \hline
\begin{tabular}[c]{@{}l@{}}Sander~\etal~\cite{sander2023dash}\\ (2023) \end{tabular} & SGX & Cloud & SGX SDK & ONNX & N & \multicolumn{1}{l|}{Y} & \multicolumn{1}{l|}{Y} & N/A & \begin{tabular}[c]{@{}l@{}}- Distributed inference\\ using garbled circuits\end{tabular} \\ \hline
\end{tabular}
}
\caption{Prior studies that leverage Confidential Computing to facilitate confidential machine learning (ConfML)}
\label{tab:confml}
\end{table*}

\subsubsection{Existing Solutions for ConfML}\label{confml}
Prior literature has focused on attaining ML using TEEs, with some aiming to overcome the aforementioned obstacles. In Table \ref{tab:confml}, we summarize and compare these past studies from various aspects.

The simplest approach is to install the whole ML training/inference process within TEEs. In such a scenario, the ML task's maximal capacity is severely constrained by the TEE's space and compute limits. Then, the aim is to optimize the efficiency of each ``bit" of TEEs' secure memory for ML computation, while striking a trade-off between the number of layers and the number of neurons in the neural network model. We outline some applicable strategies below:
\begin{enumerate}
    \item Employing inference rather than training. Training process is resource intensive, to carry out the backward propagation (\eg memory used to save model gradients and intermediate activations) \cite{zobaed2022edge}. 
    \item Selecting a modest batch size. A high batch size results in a substantial memory footprint, as each sample in the batch generates its own activations for all the neural network model layers. 
    \item Striking a balance between the feature extractor (such as convolutional layers) and the classifier (\eg fully connected layers) \cite{lawhern2018eegnet}. In fact, a correctly constructed feature extractor can reduce the feature dimension while still capturing critical information, enabling a compact classifier to perform well.

\end{enumerate}

\subsection{Application Architecture and Partitioning}
Current software systems either follow the monolithic or the micro-service-based architecture. In the former, the entire application is one tightly-coupled entity, whereas, in the latter, one application is composed of multiple loosely-coupled micro-services forming a workflow (Directed Acyclic Graph--DAG) together. While there is an extensive debate on the pros and cons of each architectural approach in terms of performance (\eg \cite{awsarticle}), there has been relatively less discussion on the impact of such architectures on confidential computing.
While monolithic applications are inflexible and often do not provide much room to boost confidential computing, the loosely-coupled and independently-developed properties of micro-service-based applications are instrumental in achieving confidential computing. 

\begin{figure} [h]
\centering
\includegraphics[width=.6\linewidth]{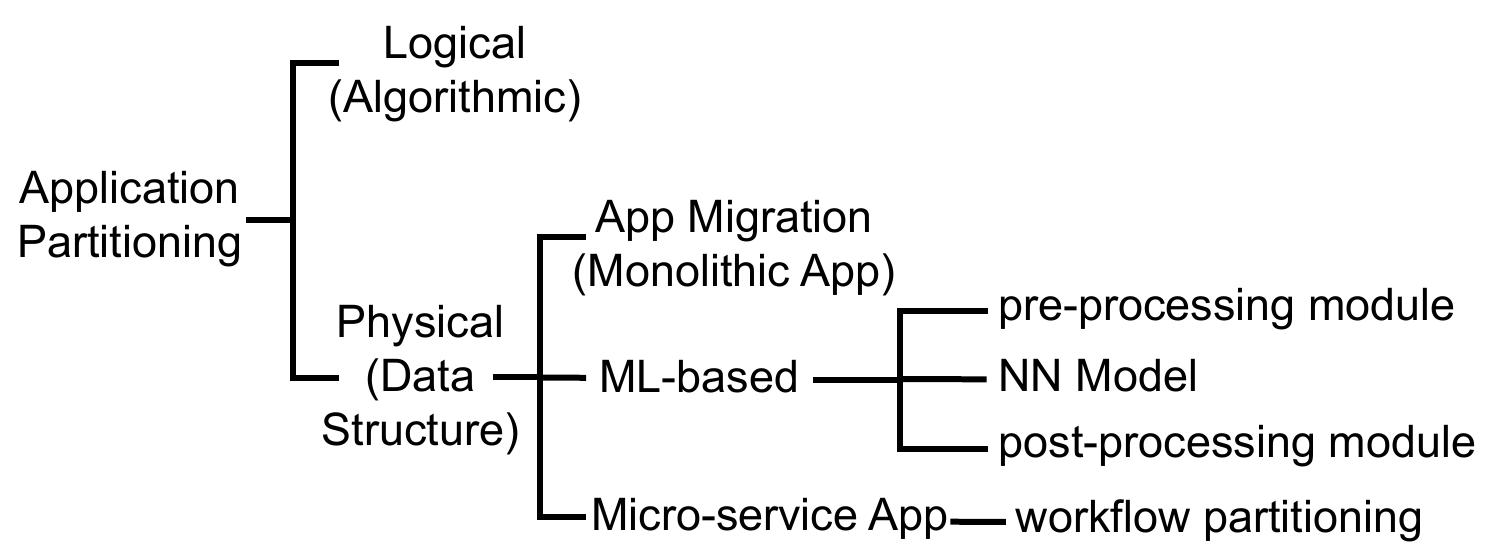}
\caption{The taxonomy of application partitioning with respect to different software architectures (microservice vs monolithic) and at various granularity levels }
\label{fig:partitionTax}
\end{figure}

The key to confidential computing at the software level is \textit{the ability to partition the sensitive and non-sensitive parts of the application and allocate them separately}. Nevertheless, as shown in Figure~\ref{fig:partitionTax}, the application partitioning itself can occur at different granularity levels.

According to the taxonomy, the application logic (\ie algorithm) can be written such that the sensitive parts are executed on the trusted machine/TEE and the non-sensitive parts are executed on the normal (non-trusted) machines. For instance, SAED~\cite{zobaed2021saed}, a tool for secure semantic search of encrypted cloud data, realizes confidential computing via partitioning the search logic/algorithm. The intelligent part that captures semantics of the user's search query is performed on-premise (on a trusted edge server). After that, the augmented search query set is encrypted and outsourced to (untrusted) cloud to perform massive ``pattern matching'' on the encrypted cloud-stored dataset. 

While the logical partitioning of the application entails developers involvement in confidential computing, there are physical partitioning approaches (see Figure~\ref{fig:partitionTax}) that leverage the \textit{application's data structures} and/or \textit{architecture} to realize confidential computing in a more transparent fashion from the developer's perspective. In this category, confidential computing can happen in three levels, namely no-partitioning; workflow-level partitioning; and ML-level partitioning. In the rest of this section, we elaborate on these levels.

\subsubsection{No-partitioning for Monolithic Applications} 
This is particularly applicable to legacy monolithic applications or those that have to run uninterruptedly, such as for monitoring industrial operations. For instance, ``measurement while drilling'' \cite{tang2021numerical} is an application that has to continuously run to avoid missing any event. Realizing confidential computing of such applications can be performed via  \emph{migration} of the application to the trusted machine. The migration itself can occur in an online (live) \cite{chanikaphon22} or offline ways.

\subsubsection{Workflow-Level (Coarse-Grained) Partitioning} 
This can realize confidential computing for applications whose architecture is based on a workflow of micro-services. For instance, fire extinguishing \cite{dunnings2018experimentally} that includes micro-services for video pre-processing, feature extraction, fire detection, and alert generation micro-services. In this case, the challenge is \emph{efficient partitioning} of micro-services (\ie tasks) of such workflows across the underlying system (\eg edge-to-cloud) such that the sensitive micro-services are executed on the trusted machines/tiers and the rest can be done on normal machines/tiers. The other challenge in this type of partitioning is how to partition the micro-service DAG so that the application can still meet its quality of service (QoS) constraints (\eg deadline) \cite{razin23}. Such partitioning can be accomplished by the underlying middleware, however, sometimes the developer or solution architects must be aware of the workflow topology.

\subsubsection{ML-Level (Fine-Grained) Partitioning} 
This approach particularly applies to ML applications that function based on neural network models that are prohibitively large for migration and/or process sensitive data. In fact, many ML-based applications are considered location-dependent, because (i) they are tightly coupled to the sensor input data (\eg captured images from a camera); (ii) there are privacy concerns in outsourcing the sensor data; and (iii) they perform inference based on large neural network (NN) models whose migration imposes a prohibitively large overhead. 

\begin{figure} [h]
\centering
\includegraphics[width=.98\linewidth]{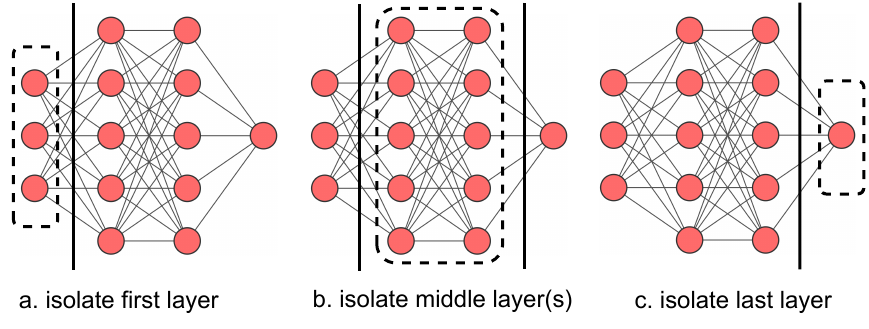}
\caption{ Partitioning neural network model of the ML applications and executing it with TEE.}
\label{fig:partitionML}
\end{figure}

For such applications, approaches based on distributed inference \cite{hosseinalipour2020multi} can be achieved that keep the source data (\eg captured images) and perform the pre-processing step on the source (trusted) machine, and then the neural network processing can be performed on a normal (potentially untrusted) machine \cite{davoodpaper}. It is even possible to vertically partition the NN model and perform some layers of it on the trusted machine and process the rest of it on the untrusted one \cite{ko2018edge}. In this manner, as shown in Figure~\ref{fig:partitionML}, the layers of an NN model are vertically partitioned and form sub-models that can be assigned to untrusted machines. The vertical model splitting is known to impose a lower data transfer overhead than an alternative method, known as horizontal splitting \cite{chinchali2018neural}. More importantly, in vertical splitting, only the intermediate feature vector has to be transferred across fogs which has two benefits: (A) reducing the amount of data-transfer overhead; and (B) preserving the sensitive  data at the trusted machine.

\subsection{Application Data Encryption}
Application Data Encryption is a vital security mechanism in trusted computing. As applications often handle sensitive user data, it is of paramount importance to secure this data not only when it is being processed, but also while it is at-rest or is in-transit.

At-rest data protection involves encoding or transforming the data before it is written to the storage, thereby, making it unreadable to unauthorized users. Even if attackers manage to breach the storage system, the encrypted data remains unintelligible without the correct decryption key, such as those provided by AES and RSA methods. In-transit application data encryption safeguards sensitive data from being intercepted and deciphered during the transmission (\eg via SSL, TLS, and HTTPS protocols).

Similarly, for in-process protection, application data encryption is critical. Upon loading the data into the secure enclave for processing, it is decrypted, and upon completion, the results are encrypted before exiting the enclave. The enclave's isolated execution and the secure storage ensure that encryption keys are safe from exposure, thereby, providing a secure environment for application data processing. Furthermore, hardware-assisted security features embedded in TEEs, such as SGX or TrustZone, provide additional layers of protection. These features, coupled with robust encryption strategies, create a secure envelope around the application data, significantly enhancing its safety and the overall security posture of the system.

\section{Remaining Challenges of Confidential Computing} \label{sec:future}
\subsection{Strengthening the Privacy Aspect of Confidential Computing}
As one of the key protective objectives of TEE, privacy must be precisely and preferably theoretically specified using information theory principles. While majority of the privacy measures are evaluated empirically through execution of the attacks, establishing privacy theoretically is challenging owing to the prevalence of assaults. 

Among contenders, \textit{differential privacy} (DP) \cite{dwork2008differential} is a widely recognized approach to achieve privacy. The DP approach ensures that the sensitive (private) information of an individual in a dataset is protected and their sensitive information is effectively anonymized, even when subjected to arbitrary side information. Importantly, the use of DP for ML-based applications has been extensively explored (\eg \cite{adnan2022federated,zhao2023federated,ponomareva2023dp,zhang2023adaptive}). However, DP has generalization challenges, which is defined as the capacity of a privacy-preserving algorithm to maintain its efficacy when applied to data beyond the initial training dataset. As an example, one may use DP-training to demonstrate the effectiveness of protection against data replication attacks via recovering the original training data from gradients generated on it. Nevertheless, it is important to note that traditional DP training is primarily designed to protect the privacy of individual records in the final model, not the intermediate gradients. Therefore, protecting against such attacks using DP might require adjusting the granularity of DP from the sample-level (\ie sensitivity of individual records in the dataset) to the feature-level (\ie sensitivity of individual features within each record). While such adjustments can theoretically enhance privacy protection against gradient-based attacks, it is also likely to significantly impact the model's performance. In fact, there is a trade-off between privacy and utility in such protections.

\subsection{Anonymous Attestation}
Remote attestation needs to be further extended to anonymous attestation. Remote attestation verifies to a third-party that a TEE is being executed on a machine without its integrity being compromised, whereas, anonymous attestation is an addition to the remote attestation that ensures the anonymity of the TEE user. Thus, it just reveals that a TEE is employed, but not which one. 

A straightforward approach can be employing asymmetric encryption in which each TEE utilizes the same secret key. Using the secret key, the TEE may authenticate itself to a third party. Because the secret key is identical, the verifier is incapable of identifying the TEE. However, this approach has significant downsides. If even a single TEE is hacked or the secret key is compromised, the attestation mechanism can be rendered ineffective. This is an additional significant problem for anonymous attestation techniques, which must be able to detect revoked customers without knowing who they are. That is why devising an anonymous attestation mechanism is the need of an hour to maintain the integrity of \cc. 

\subsection{Addressing Vulnerabilities of TEE: SGX, TrustZone, and SEV}
Developing strong and effective \cc\; solutions must take into account the constraints posed by the usage of TEE. The problem is that side-channel and reverse-engineering attacks are not included in the threat model of the TEEs, such as Intel SGX. Moreover, there are a number of security alerts reported concerning TrustZone, including kernel and driver issues, as well as hardware-related vulnerabilities that affect various hardware components of the platform~\cite{pinto2019demystifying}. 

Software developers may think that TEE is absolutely safe, but this is not the case. They must take into account defects and vulnerabilities in hardware and software components of TEEs. It is important for future research to address vulnerabilities of the current TEE technologies such as Intel SGX, ARM TrustZone, and AMD SEV. While these technologies offer strong data security measures, they are not immune to attacks. For example, the threat model of Intel SGX does not consider side-channel and reverse-engineering attacks, which are potential hardware vulnerabilities. Additionally, there have been security alerts concerning ARM TrustZone, such as kernel and driver issues, as well as hardware-related vulnerabilities that affect various hardware components of the platform. 

As such, it is important for future research to focus on finding solutions for these vulnerabilities and ensure the maximum achievable TEE securities. This can be achieved via identifying the specific weaknesses of the TEE technologies and developing new and more effective solutions to address them.

\subsection{Integration of Distributed Trust and Confidential Computing}
Distributed trust \cite{wei2022trust} is defined as trust models and solutions where the trust is established as a result of consensus across multiple nodes/entities. Blockchain \cite{krichen2022blockchain} and distributed ledger \cite{burkhardt2018distributed} are popular examples of distributed trust. Solutions based on distributed trust have the potential to be integrated with and enhance the security and scalability of confidential computing. For instance, the decentralized nature of blockchain can provide an immutable and tamper-proof record of all data and computations, while confidential computing can protect the data and computations themselves from unauthorized access and tampering. Additionally, distributed trust can provide secure multi-party computation, enabling multiple parties to securely and privately collaborate on sensitive data and computations, while preserving the privacy and security of the underlying data.

While traditional trust models often rely on centralized trust authorities, which can become bottlenecks and introduce vulnerabilities, distributing trust across multiple entities can mitigate the risk of single points of failure and more resilient protection of sensitive data. To further advance the field of confidential computing, however, the following aspects of distributed trust have to be further explored:
\begin{itemize}

\item \textit{Trust establishment and management} Developing novel mechanisms to establish and maintain trust across heterogeneous and dynamic environments, such as edge-to-cloud continuum. Techniques for trust negotiation, delegation, and revocation can also be investigated to ensure secure and efficient collaboration across various participating entities.

\item \textit{Zero Trust Access in IoTs} Zero Trust model is a security concept centered on the belief that organizations should not automatically trust anything inside or outside their perimeters rather must verify everything trying to connect to their systems before granting access. This concept is particularly relevant to IoT, where devices are often highly diverse and can present significant security vulnerabilities. By adopting zero trust protocols, organizations can ensure that each IoT device is authenticated and its behavior continuously monitored, regardless of its location or network. Zero trust can be particularly effective when combined with other security technologies such as micro-segmentation and network access control (NAC), ensuring that IoT devices have only the minimum necessary access rights, and that any suspicious behavior can be rapidly detected and addressed. Implementing Zero trust in IoT environments can therefore greatly enhance their resilience against potential security threats.

\item \textit{Consensus algorithms} Designing new consensus algorithms that can ensure trust and integrity in distributed confidential computing systems, while maintaining efficiency and scalability. These algorithms should be robust to malicious activities and adaptable to changing network conditions.

\item \textit{Secure data sharing and collaboration} Implementing secure data sharing mechanisms that allow multiple parties to collaboratively process sensitive data while preserving data privacy and confidentiality. This can involve leveraging techniques such as multi-party computation \cite{riazi2018chameleon}, federated learning \cite{zhao2023federated,adnan2022federated}, and blockchain-based solutions \cite{wan2019blockchain}.

\item \textit{Monitoring and auditing} Developing monitoring and auditing tools to ensure the security and compliance of distributed trust deployments. These tools should be capable of detecting and mitigating security threats, as well as providing transparency and accountability to all participating entities.

\item \textit{Performance optimization} Investigating methods to optimize the performance of distributed trust deployments while maintaining security and privacy guarantees. This can involve exploring trade-offs between security and performance, as well as developing adaptive algorithms that can dynamically adjust to different workloads and network conditions.
\end{itemize}

\subsection{Dedicated TEE Designs for General ML}
Efficient ML execution requires modern computer architectures with parallelization features, such as multi-threading on
CPUs and accelerators (\eg GPU, TPU, etc.) to perform matrix multiplication in forward and backward passes. These features are achieved via application-specific hardware (a.k.a. ASICs)
that most of the current TEEs do not have. One solution is to equip TEEs with such parallelization features to increase their computational capability for ML-based applications. However, such new features increase the TCB size. Moreover, synchronization bugs can cause severe vulnerability
of SGX~\cite{cloosters2020teerex}, and one can also expect a considerable number of bugs that potentially exist in GPU-TEEs. 

Avoiding such a dilemma depends on how to limit the trust boundary and reduce the TCB when applying parallel processing. One approach is similar to the way TPM operates: one ML accelerator (GPU/TPU) is physically isolated from the rest of the motherboard and the processing system. The accelerator will be accessed solely through a secure bus from the TEE-enabled CPUs. Although this approach constrains the trust boundary and no one can physically breach the GPU, there are also other possible approaches such as ML as a Service (MLaaS) \cite{mo2022sok} with proper remote attestation and verification.


\section{Summary}\label{sec:conclsn}
The overarching goal of \cc\;is to establish end-to-end data security and privacy that includes data at-rest, data in-transit, and data in-use across various computing systems. While other types of security has been extensively explored, \cc\;has mainly emerged to deal with the security of data in-use (while being processed). Confidential computing has become a vital research area due to: \textbf{(a)} the exponentially increasing volume of data generated by various sources, ranging from IoT-based sensors to social media activities; \textbf{(b)} data processing commonly occurs off-premises and on third-party servers---across edge-to-cloud continuum---where the need for secure processing of sensitive data is of paramount importance; and \textbf{(c)} prevalence of data-driven ML applications that process and identify confidential information about businesses and individuals. 

Accordingly, this study aims at providing an overarching understanding of the fundamental concepts and of confidential computing at the hardware, middleware, and application levels. Moreover, this study surveys the existing solutions and state-of-the-art techniques available for \cc. More specifically, we delved into the core components of confidential computing, such as Trusted Execution Environments (TEEs), secure enclaves, and examined their applications in diverse domains, including cloud computing, IoT, edge computing, and particularly with respect to ML applications. In addition, we discussed the importance of establishing trust in both hardware, middleware, and software levels, along with the role of remote attestation procedures and trusted application development frameworks in achieving \cc.

Despite the considerable progress made in the field of confidential computing, several challenges and research opportunities have remained unexplored. Future research endeavors should focus on enhancing privacy aspects within confidential computing solutions, developing more secure and anonymous attestation mechanisms, and addressing vulnerabilities in current trusted execution environments. Furthermore, it is crucial to explore scalable and efficient \cc\;approaches that can effectively handle the massive volume of sensitive data being processed via ML-based solutions.

In conclusion, this survey offers a thorough overview of the confidential computing landscape, highlighting its importance in the ever-evolving digital age. By examining existing solutions, challenges, and future research directions, we hope to inspire researchers and practitioners alike to continue advancing this field, ultimately ensuring true secure processing of sensitive data across various applications and domains.

\section*{Acknowledgments}
 We would like to thank the anonymous reviewers of the paper. This research is supported by the National Science Foundation (NSF) under awards\# CNS-2047144, CNS-2117785, and CNS-2007209.

\bibliographystyle{vancouver-modified} 
 \bibliography{references}


\end{document}